\documentclass[manuscript,screen]{acmart}
\usepackage{enumerate}
\usepackage{comment}

\AtBeginDocument{%
  \providecommand\BibTeX{{%
    \normalfont B\kern-0.5em{\scshape i\kern-0.25em b}\kern-0.8em\TeX}}}

\setcopyright{acmcopyright}
\copyrightyear{2018}
\acmYear{2018}
\acmDOI{10.1145/1122445.1122456}

\begin{document}

\title{Advances in Process Optimization: A Comprehensive Survey of Process Mining, Predictive Process Monitoring, and Process-Aware Recommender Systems}


\author{Asjad Khan}
\affiliation{%
 \institution{University of Wollongong}
 \streetaddress{Northfields Ave}
 \city{Wollongong}
 \state{NSW}
 \country{Australia}}

\author{Aditya Ghose}
\affiliation{%
 \institution{University of Wollongong}
 \streetaddress{Northfields Ave}
 \city{Wollongong}
 \state{NSW}
  \country{Australia}}

\author{Hoa Dam}
\affiliation{%
 \institution{University of Wollongong}
 \streetaddress{Northfields Ave}
 \city{Wollongong}
 \state{NSW}
  \country{Australia}}
\email{asjad@outlook.com}

\author{Arsal Syed}
\affiliation{%
 \institution{University of Nevada}
 \streetaddress{1664 N Virginia St, Reno}
 \city{Las Vegas}
 \state{Nevada}
  \country{U.S.A}}
\email{syeda3@unlv.nevada.edu}

%\renewcommand{\shortauthors}{Asjad Khan et al.}

\begin{abstract}

Process analytics approaches allow organizations to support the practice of Business Process Management and continuous improvement by leveraging all process-related data to extract knowledge, improve process performance and support decision making across the organization. Process execution data once collected will contain hidden insights  and actionable knowledge that are of considerable business value enabling firms  to take a data-driven approach for identifying performance bottlenecks, reducing costs, extracting insights and optimizing the utilization of available resources. Understanding the properties of ‘current deployed process’ (whose execution trace is often available in these logs),  is critical to understanding the variation across the process instances, root-causes of inefficiencies and determining the areas for investing improvement efforts. In this survey we discuss various methods that allows organizations to understand the behaviour of their processes, monitor currently running process instances, predict the future behavior of those instances and provide better support for operational decision-making across the organization.

%Process mining techniques can leverage this execution data to mine actionable knowledge, discover insights about performance bottlenecks, frequent defects, their root causes and other sources of inefficiencies. 

\end{abstract}


%\ccsdesc[500]{Computer systems organization~Embedded systems}


\maketitle

\section{Introduction}

%Beheshti et. al. \cite{benatallah2016process} have described process analytics as \textit{ ``Business analytics is the family of methods and tools that can be applied to process execution data in order to support decision-making in organizations by analyzing the behavior of completed processes (i.e., process controlling), evaluating currently running process instances (i.e., business activity monitoring), or predicting the behavior of process instances in the future (i.e., process intelligence).} 

%By doing so, it aims to make businesses to stay competitive by for example improving the Quality of Service and efficiency of Processes and providing the relevant decision support to the knowlege workers invovled in the exuction of those processes. 
%This involves extracting useful insights from event logs and other process related data to develop various capabilities such as:

Business processes are at the heart of every organization in the emerging knowledge economy. Process execution data contains hidden insights and actionable knowledge of considerable business value. Process analytic approaches allow organizations to support the practice of Business Process Management and continuous improvement by leveraging all process-related data to identify performance bottlenecks, reducing costs, extracting insights and optimizing the utilization of available resources. Over the past two decades, the process mining research community has primarily focused on investigating problems such as (automated)  process discovery, process conformance checking and process enhancement. Within the broad umbrella of {\em process mining}, several methods have been proposed to provide support during the process redesign and process analysis/diagnosis phases of the BPM life-cycle. e.g constructing simulation models from process logs, performing organizational or social-network mining, case outcome predictions and so on \cite{van2011process}. These methods allow organizations to leverage process execution logs to answer questions like; \emph{Does the process behave as expected?} or \emph{Are there any bottlenecks that negatively impact process performance?} and \emph{Do the logged instances conform to applicable laws and regulations?} \cite{dumas2013business}. While process mining topics like discovery, conformance checking, and enhancement continue to be active areas of research, several newer methods have grown in prominence that leverage a much more diverse range of data sources and offer a much more sophisticated set of capabilities. 

Process analytics allows organizations to support the practice of Business Process Management by leveraging all process-related data to extract knowledge, improve process performance and support managerial-decision making across the organization\cite{dumas2018business}. Process analytics involves a sophisticated layer of data analytics built over the traditional notion of process mining. While process mining addresses the problem of reverse engineering process designs from process execution data (process logs), process analytics extends the scope and addresses the more general problem of leveraging data generated by, or associated with, process execution to obtain actionable insights about business processes. Process analytics leverages a range of data, including, but not limited to process logs, event logs, provisioning logs, decision logs and process context. Process analytics can be used to obtain predictive insights (how will a given process work out?), diagnostic insights (why did a process generate a given outcome?) as well as prescriptive insights (what should be done next in a process?). Process data can for example be used to prescribe what resources should be allocated to process tasks, or predict whether a process will achieve its goals.

Process analytics enables firms to take a data-driven approach for identifying performance bottlenecks, reducing costs, extracting insights and optimizing the utilization of available resources.It encompasses methods, tools and techniques that allow firms to understand the behaviour of their processes, monitor currently running process instances, predict the future behaviour of those instances and provide better support for operational decision-making across the organization \cite{benatallah2016process}. Process analytics can also be viewed as an organizational capability that enables firms to stay competitive by better understanding their business processes and identifying areas of improvement. Some examples of these capabilities include: 

\begin{itemize}
    \item Mining the logs of historical (completed) traces to better understand the process behaviour, retrospectively. 
    
    \item methods to analyse processes through data to gain insights about performance bottlenecks and identifying the root causes of undesired process behaviour/outcomes to support evidence-based process analysis and improvement 
    
    \item Understanding resource behaviour and optimizing the utilization of available resources for future instances. 
   
    \item  Supporting operational decision-making across the organization during process execution by for example, making process behaviour forecasts for running cases. 
    
    \item Providing decision support to knowledge workers involved in the execution of knowledge-intensive, unstructured or semi-structured processes.
    
    \item Supporting the practice of risk management by assisting in early identification and possible mitigation of undesired effects.
    
    %\item Analysing performance of processes to gain actionable knowledge,  and and uncovering root-causes 
    
    %\item Develop predictive monitoring capabilities where for example we predict the future behavior of those instances and provide better support for operational decision-making across the organization. e.g Predict the \textit{cycle time, resource usage and cost} of a given process instance ? 
\end{itemize}

%We can broadly classify these contemporary areas as process behaviour analysis or  behaviour prediction and  supporting process related decisions\ag{(this line will be revisited after taxonomy is decided)}. 

%the form of task allocation and resource allocation recommendations that help achieve optimal performance and assist in possible mitigation of undesired effects

In this work, we aim to cover process analytic techniques that have received relatively less attention from the research community and therefore offer a fertile ground for exploration and contributions. We can broadly classify them into three major themes: \textit{Mining Process Behaviour, Predictive Process Monitoring, and Process Decision Support}. In mining process behaviour, we focus on techniques that can help us analyze historical (completed) traces to better understand the process behaviour, retrospectively \cite{van2011process}. Process Mining methods are used for post-mortem analysis of completed business processes to assist with process improvement and redesign efforts. They can help in optimizing process performance by mining process behaviour, identifying bottlenecks (or various sources of inefficiencies and frequent defects), and understanding their root causes. Predictive Process Monitoring methods on the other hand aim at analyzing execution data (e.g. event logs) of a business process at run-time to forecast the future state of the executions of a business process \cite{teinemaa2019outcome}. These methods are an effective decision support tool for continuously monitoring performance of process instances and reducing the overall risk associated with negative outcomes. Lastly, Assisting organizations in operational decision support remains one of the major goals of process analytics and process management systems. We discuss the topic of decision support, specifically in the context of: \textit{(i)} supporting knowledge-intensive processes by recommending the best alternative given a decision-making point during process execution. \textit{(ii)} providing intelligent assistance for resource allocation and (\textit{iii)} reducing risks associated with operational and strategic decisions for processes operating in dynamic environments. Overall, We see these capabilities and themes in combination represent an important area for further exploration and research.

%We can also view this past process execution data as organizational experience representing either positive (value-adding) outcomes or negative (non-optimal) outcomes.  
%Some of these topics under these themes have been well investigated in the current literature over the last decade(under various titles), while others have received little attention, offering 
%In particular, these methods can assist organizations in 

%need to be revisited by the community as a major thread of research [ALT: ...

% Overall, this paper aims at providing a comprehensive survey of process analytic methods and capabilities with a focus on various emerging topics that will be relevant to current and future researchers. Our specific contributions are as follows: \textit{(i)} We present a systematic literature review of key topics identified under the above mentioned themes in order to develop a unifying view of the field. Our survey consists of both popular and under-investigated topics relevant for the current and future practice of Business Process Management. \textit{(ii)} We provide a taxonomy for the identified topics and identify relevant research gaps along with potential research directions. 

%\textit{(v)} provide a research roadmap for future research. 
 
The rest of the paper is structured as follows: Section 2 reviews basic concepts of BPM, and discusses the relationship between Process Analytics and Business Process Management. Section 3 describes the search protocol used for conducting the systematic literature review. Section 4 discusses methods for understanding process behaviour using  logs of historical (completed)
traces. Section 5 surveys the selected studies related to predictive process monitoring methods and provides a taxonomy to classify them. In Section 6, we investigate the theme of decision support in the context of structured, unstructured and semi-structured knowledge intensive processes.

\section{PRELIMINARY CONCEPTS}

%\textbf{Business Processes:}

%Business processes are key instruments in organizing activities whose outcome is a product or a service \cite{di2015knowledge}. Dumas et al. \cite{dumas2013business}  have defined business processes as \textit{``Events, activities and decisions involving multiple actors and resources, that collectively lead to an outcome that is of value to an organisation or its customers''.} Alternatively, we can define a process as: \textit{``A business process is a collection of inter-related events, activities, and decision points that involve a number of actors and objects, which collectively lead to an outcome that is of value to at least one customer.''} \cite{dumas2018fundamentals}. \\

\noindent
\textbf{Business Process Management:} 

%Several business management approaches like Continuous Process Improvement (CPI), Business Process Improvement (BPI), lean management and six sigma have been proposed over the years to measure and analyze the quality and efficiency of processes \cite{van2011process}. BPM borrows ideas from many of these disciplines and offers a holistic management discipline centered around some core elements like strategic alignment, governance, people and culture \cite{rosemann2015six}.  

Business Process Management (BPM) follows a paradigm of ``process thinking'' where we take a process-centered approach towards understanding and improving business processes  \cite{vom2010handbook}. Business processes are key instruments in organizing activities whose outcome is a product or a service \cite{di2015knowledge} Processes are described by \textit{`Events, activities and decisions involving multiple actors and resources, that collectively lead to an outcome that is of value to an organisation or its customers'} \cite{dumas2013business}. BPM is defined as:\textit{"a body of principles, methods and tools to design, analyse, execute and monitor business processes, with the aim of improving their performance."} Effective management of enterprise-wide processes enables organizations to control the process outcomes and focus their efforts on identifying and improving areas of high impact. BPM is concerned with management, compliance, and the redesign of business processes. It aims at bringing predictability, transparency and consistency to business operations \cite{dumas2013business}. BPM practice is supported by various modeling tools and analysis methods that have been proposed over the years. Such methods and tools enable continuous performance improvement by optimally allocating scarce resources and tracing performance bottlenecks as possible examples. 

%Similarly, it helps organizations in innovate by identifying demands and opportunities that deliver enhanced value to the process customer \cite{dumas2013business}. 

%It also helps ensure that organizational strategic goals and criteria are aligned with organizational capabilities \cite{dumas2013business}. 

%Value-Driven Business Process management framework \cite{franz2011value} focuses on achieving desired outcomes by prioritizing transparency about organizational operations and streamlining organizational processes. x

\noindent
\textbf{Process Management Systems:} Formerly known as workflow management systems, Process Management Systems (PMS) are designed to facilitate the operations of an organization by supporting all the phases in the business process life cycle (e.g via managing the process routing and allocation of resources) \cite{di2015knowledge}.  In doing so, a PMS also records event data representing the execution trace of various deployed processes. Currently, there exist various types of Business Process Management systems that can be distinguished based on \emph{degree of support'} they provide and `orientation on process data' \cite{dumas2013business}. Examples include groupware systems, ad-hoc workflow systems, production workflow systems and case handling systems \cite{ouyang2009business}.\\\\

\noindent
\textbf{Process Data}

Processes deployed inside and across an enterprise, when executed can leave an operational data footprint in the form of an event logs, which if analyzed can be a valuable source of insights to support the management and improvement of business processes. This Process execution data can transformed into an event log describing the sequence of activities that were performed, along with the resources involved in the execution. Each entry of the process execution log(also known as event logs) represents an \emph{event}, which records the occurrence of an activity at a particular point in time and belongs to precisely one case (case represents a unique process instance). Events can be characterised by multiple descriptors (attributes), including an event (or activity name), a unique case identifier (e.g case ID), a timestamp and optionally details of resource responsible for executing the task of the process instance. An \textit{event} refers to an activity (or a step) in the process and belongs to a process instance or a case \cite{dumas2013business}\cite{weske2019business}. The sequence of ordered events within a case form a trace. We use the following definition to define an event:  
%Formally, an event is a tuple $\left(act, case_id, time,trans_type, resource, \left(d_{1}, v_{1}\right), \ldots,\left(d_{m}, v_{m}\right)\right)$ where $act$ is the activity name representing a step in the process $case_id$ represents a specific process instance, $time$ is the timestamp at which the event took place, $trans_type$ represents the state when event was recorded, $resource$ refers to the resource involved in the execution of a specific activity and $\left(d_{1}, v_{1}\right) \ldots,\left(d_{m}, v_{m}\right)$ (where $m \geq 0$ ) are the optional event-specific (or case-specific) attributes and their values. Formally 

\textbf{Definition \cite{van2016process}: } An event $e$ is tuple $e=(c, a, t, r) \in$ $\mathcal{U}_{\text {case }} \times \mathcal{U}_{\text {act }} \times \mathcal{U}_{\text {time }} \times \mathcal{U}_{\text {res }}$ referring to case $c$, activity a, timestamp $t$, and resource $r$ of event e. An event $\log L$ is a multiset of events, i.e., $L \in \mathcal{B}\left(\mathcal{U}_{\text {case }} \times \mathcal{U}_{\text {act }} \times \mathcal{U}_{\text {time }} \times \mathcal{U}_{\text {res }}\right)$.

%An event log may carry additional event-specific attributes, naturally ordered by the associated timestamps to represent a sequence of events. 

%including an event class (or activity name), a resource, and a timestamp, associated with a process instance or case . 
%We assume that the events are discrete and drawn from a finite set. 
%Resources associated with an activity can be either continuous (e.g., time to complete a sub-task), or discrete (e.g., type of skills needed). 
% The universe of all events is hereby denoted by $\mathcal{E}$ 

Apart from event logs (which record sequential events representing activities in a case instance) organizations can leverage a diverse set of heterogeneous data sources to gain a holistic view of all process-related aspects. Various sources of structured and unstructured process-related execution data are available but often times are scattered across several systems. Examples include: 

\noindent
\begin{itemize}
  \item Enterprise Resource Planning Systems
  \item Contextual data e.g from real-time IOT sensors 
  \item Web data. e.g Social Media Sentiment
  \item Provisioning Logs and decision logs
  \item Unstructured and Semi-Structured data 
\end{itemize}

\noindent
\textbf{BPM lifecycle:}

Heiskanen et al. \cite{heiskanen1997bridging} have described the BPM lifecycle as \textit{"The entry point to the cycle is the design and analysis phase, where the business processes are identified and provided with a formal representation. Newly created models and models from past iterations are verified and validated against current process requirements. In the configuration phase, the systems to use are selected, and the business processes identified before are implemented, tested, and deployed. During the enactment phase, the processes are operated, and the process execution is monitored and maintained. The resulting execution data is processed by the techniques of the evaluation phase, for example process mining. Using the knowledge gained from one iteration, the next iteration can be started by redesigning the business processes."} We briefly describe the steps below: 

\begin{itemize}

    \item \textbf{Process Identification/(Re)design}: 
    In process identification, after requirement analysis, process models are designed using a suitable modeling language. Existing models can be further refined/improved based on insights gathered in the previous cycle.  
    
    \item \textbf{Process Analysis/diagnosis}: 
    In this step, process logs are typically analysed to diagnose problems and identify areas of potential improvement. Process Discovery methods are also useful here, allowing the analyst to reverse engineer the models from recorded process execution logs \cite{di2012knowledge}. 
   
    \item \textbf{Process Implementation}: In Process implementation (also known as process enactment), a process management engine is sometimes used to support the process enactment by instantiating a process instance where tasks are assigned to the relevant resources for execution. A PMS can also manage process routing(control-flow) by considering which tasks are enabled for execution \cite{benatallah2016process}.  
    
    \item \textbf{Process Monitoring and Controlling}: 
    Organizational process are monitored by the various process support and management tools while executed tasks are tracked and recorded to generate execution traces for later analysis. 
    
\end{itemize}

\textbf{Business Process Paradigms:} 

It is useful to characterise various types of processes based on the degree of structuring and predictability they exhibit. Process paradigms describe work activities that a process management (or process-support) system can handle \cite{benatallah2016process}. Traditional process management systems have provided support for modeling, monitoring and management of structured processes. However, business processes often contain reusable patterns with elements of unpredictable nature, requiring a certain degree of flexibility. Ciccio et al. \cite{di2015knowledge} have described the spectrum of process management as follows:  

\begin{itemize}

  \item \textbf{Structured processes}: Structured processes represent predictable routine work (e.g administrative processes). They have a predefined schema (defined apriori in a process model) where process logic and ordering of activities, their dependencies and allocation of resources is known in advance. 
  
  \item  \textbf{Structured processes with ad-hoc exceptions}: They allow a certain degree of flexibility as unanticipated exceptions can cause deviations during the execution of a process instance. Some of these deviations can also be anticipated in advance and incorporated into the process design via exception handlers.
  
  \item  \textbf{Unstructured processes with predefined segments}: Here structured process fragments can be pre-defined on a per-case basis(based on policies and regulations) but overall process logic cannot be specified in advance.
  
  \item  \textbf{Loosely structured processes}: Loosely structured processes require adaptation strategies, as the ordering of activities is hard to anticipate in advance, making the overall process structure less rigid. Here constraints (described by policies and business rules) are defined ahead of time to prohibit undesirable behaviour.  
  
  \item  \textbf{Unstructured processes}: Unstructured processes are often knowledge-centric (dependent on human judgement and expertise) and represent complex, non-routine business process. They are collaborative in nature, driven by rules and dynamic events where the structure of a process evolves based on the operational context.
 
\end{itemize}

Readers can refer to \cite{kemsley2011changing} \cite{harrison2018human} \cite{rosenfeld2011bpm} for a more detailed discussion on the process spectrum.

\section{Research Methodology}

We identified and classified the most relevant process analytics studies by conducting a \textit{systematic literature review (SLR)} according to the scientifically rigorous guidelines described in \cite{kitchenham2004procedures}. First, we formulated a list of research questions representing our research goals. Next, guided by these questions, we describe the relevant search strings used for querying a database of academic papers. We then applied inclusion/exclusion criteria to the retrieved studies and filtered out the irrelevant ones. Finally we divided all relevant studies into primary and subsumed ones based on their contribution.

\subsection{Research Questions}

Our paper aims to conduct a systematic literature review by analysing research studies related to process analytics. Our goal is to understand the recent developments in the field of process analytics by examining the following research questions:

\begin{enumerate}[{}]

    \item RQ1  What is the body of recent and relevant academic publications within the field of business process analytics?
    \item RQ2   What family of methods exists for diagnostic analytics that lets us mine relevant process data in a retrospective (post-mortem) fashion?
    \item RQ3  What aspects of business processes can be predicted?
    \item RQ4   What are some of the state-of-art methods for predictive monitoring tasks? 
    %How were the proposed methods evaluated and what is the relative performance of these methods? (Note to self: This question is optional)
    \item RQ5   What role can prescriptive process analytic methods play in automating or supporting decision making in the context of business process execution?
    \item RQ6   How do we characterise these findings in a taxonomy?  
    \item RQ7  What should be the future research focus of BPM and process analytics?
\end{enumerate}

RQ1 is the core question that identities existing methods to support the practice of business process management. It allows us to identify a set of classification criteria. Given the richness of the process analytics literature, this overarching question is then decomposed into subsequent research questions to have a well-delimited and manageable scope. RQ2 aims to identify methods that attempt to understand the process behaviour using offline data in order to diagnose possible performance problems and identify areas of potential improvements. Given the vast number of publications in the broader field of process mining, we focus on contemporary topics that have not been examined in great detail so far. RQ3 and RQ4 investigate aspects of business processes that can be predicted by means of machine learning techniques (also known as business process monitoring techniques). RQ5 explores prescriptive methods with the potential of providing decision support both at strategic and operational level.  

We categorize our findings based on input data required, type of algorithm employed, validation method, and tool support. (Note to self: The strategy for taxonomy isnt finalized yet). 

\subsection{Study Retrieval} 

Keeping the research goals in mind,  and following the guidance given in \cite{kitchenham2004procedures} we define the search string. We drive a set of relevant keywords from our subject matter expertise: 

\begin{itemize}

\item[] ``business process'' — generic term for retrieving most process analytics papers.
\item[] ``mining'' - Here a relevant study must take as input a event dataset and proposes a technique for analysing, extracting actionable knowledge. 
\item[] ``diagnostic'' — diagnostic methods perform post-mortem analysis of process data in an attempt to understand what happened in the past and to analyse the root causes of performance issues. 
\item[] ``monitoring`` — monitoring methods are concerned with run-time predictions of process outcomes.
\item[] ``prediction'' — a relevant study that estimates or predicts various aspects/properties of a business process.
\item[] ``prescriptive" - studies that are concerned with action recommendations
\item[] ``decision-support" - Methods that provide decision support for achieving process goals.
\item[] ``recommendations" - Methods that provide recommendations for decision support.

\end{itemize}

\noindent   
Based on these selected keywords and criteria, we constructed the following search phrases: 

\begin{itemize}

\item ``process mining AND performance improvement''
\item ``process mining AND resource management''
\item ``business process prediction''
\item ``predictive business process monitoring''
\item ``prescriptive process analytics AND decision Support''
\item ``prescriptive process analytics''
\item ``process recommendations AND decision-support''
\item  ``decision support AND process analytics''
\item  ``decision support AND knowledge-intensive processes''
\end{itemize}

\subsection{Study Selection}

We then picked the Google Scholar database \cite{gusenbauer2019google} as our primary search engine for retrieving relevant studies based on our final search strings. We double-checked the retrieved studies with six other literature search sources, including SpringerLink, Scopus, IEEE Xplore, ScienceDirect and ACM Digital Library. These major database electronic literature databases cover the scientific publications within the field of computer science.  After search was conducted(in August 2020), it returned more than 5000 papers. 

\noindent  
\textbf{\textit{Exclusion criteria:}} Non-English and Duplicate studies (appearing in multiple databases) were removed. Short Papers with page length 6 and workshop papers were also excluded. Studies which propose method whose input is not process data are also excluded. Lastly, studies where the main contribution of the paper is a case study or a tool implementation instead of novel model with proper evaluation are excluded.  

\noindent  
\textbf{\textit{Inclusion criteria:}} Based on the meta-data (titles, abstract) of remaining papers, we then filtered studies that appeared out of scope based on inclusion criteria(as suggested by \cite{okoli2015guide} \cite{fink2019conducting}).

\textbf{IN1:} We picked the study only if: (i) it is concerned with \textit{analysis and mining process data}. (ii) proposes techniques for \textit{predictive monitoring} of processes. (iii) investigates methods related to \textit{prescriptive analytics} and \textit{decision support} in the context of \textit{structured, unstructured and knowledge-intensive business processes}. 

\textbf{IN2:}  Study is published in 2011 and later were all included, even if they had fewer than 5 citations. However, for papers published before 2011 we used the snowballing technique. 

\textbf{IN3:} The study clearly defined \textit{research context, goals} and proposes a \textit{novel method} that has been properly evaluated with sound experiments. 

\textbf{IN4:} The study is \textit{peer-reviewed} and published in a reputable venue. 

For final inclusion, a given study must meet the above inclusion criteria. We applied the inclusion criteria IN2 by configuring the search engine's filter settings. For applying the IN1 and IN3 criterion,  was assessed by reading title, abstract and skimming the relevant sections of the paper. After filtering the search results by applying inclusion and exclusion criteria, we identified primary and subsumed studies. Primary studies constitute an original contribution to the field and subsumed are ones that do not substantially improvement with respect to the original contribution. The application of the exclusion criteria resulted in 734 relevant studies out of 1319 works selected in the previous step. 

\subsection{Taxonomy}

We categorize the selected works using different dimensions specifying the typology of the existing methods and their characteristics. 

\begin{figure}[htp]
    \centering
   \includegraphics[scale=0.35]{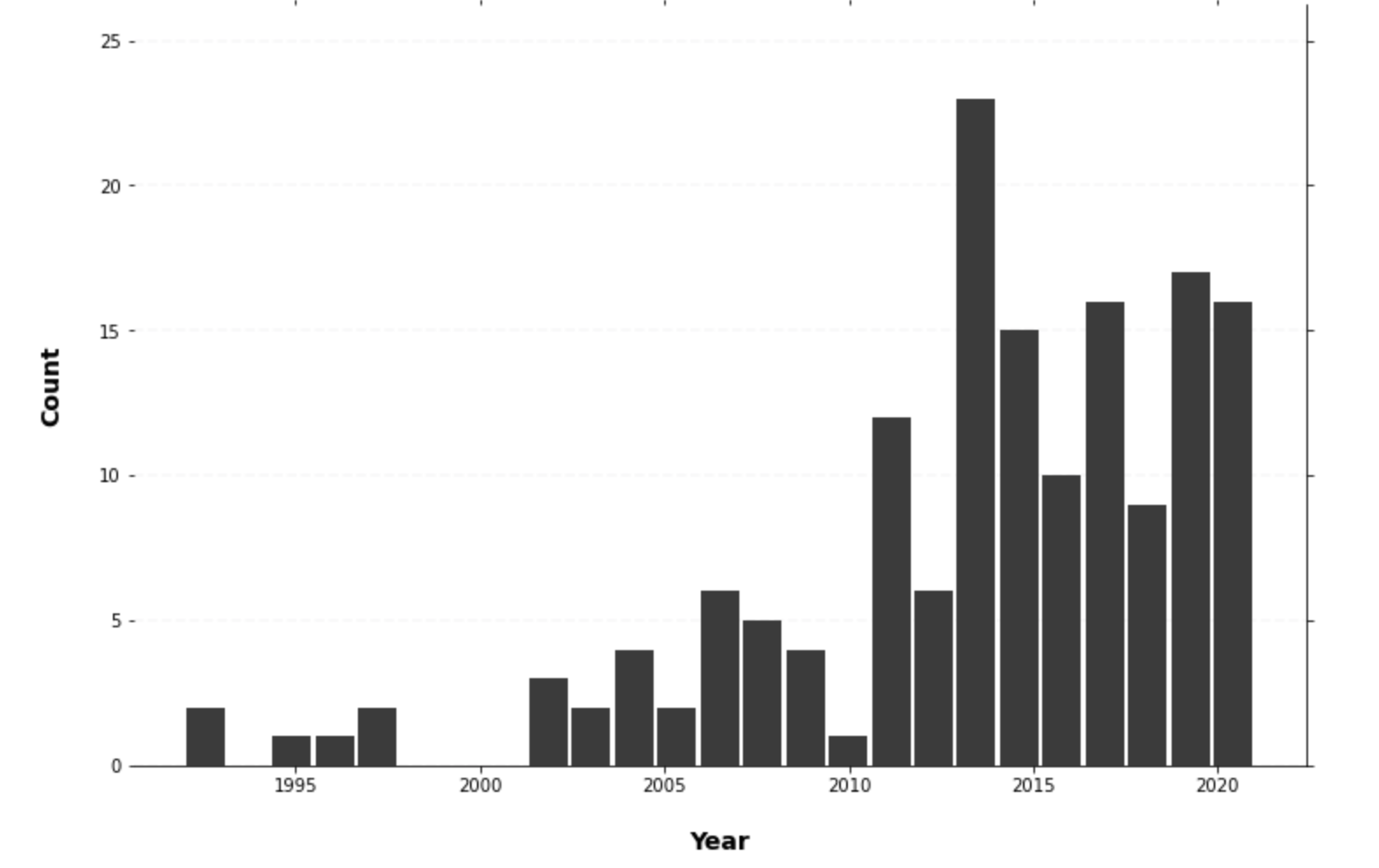}
    \caption{Publications dates of included literature}
\end{figure}

In particular, each study can be decomposed and organized along the following dimensions: \\
- Input data \\
- Outcome \\
- Process perspective (control flow, resources, data) \\
- Family of algorithms(the main algorithm used in the study) \\ 
- Evaluation data (real-life or artificial logs) and application domain (e.g., insurance, banking, healthcare) \\ 
- Implementation (standalone or plug-in, and tool accessibility) 

Lastly, we also observed that currently, many of the same mining problems are being studied under different names, and proposed methods are being evaluated in an ad-hoc manner with no common experimental setups or evaluation measures.

\begin{figure}[htp]
    \centering
    \includegraphics[scale=0.50]{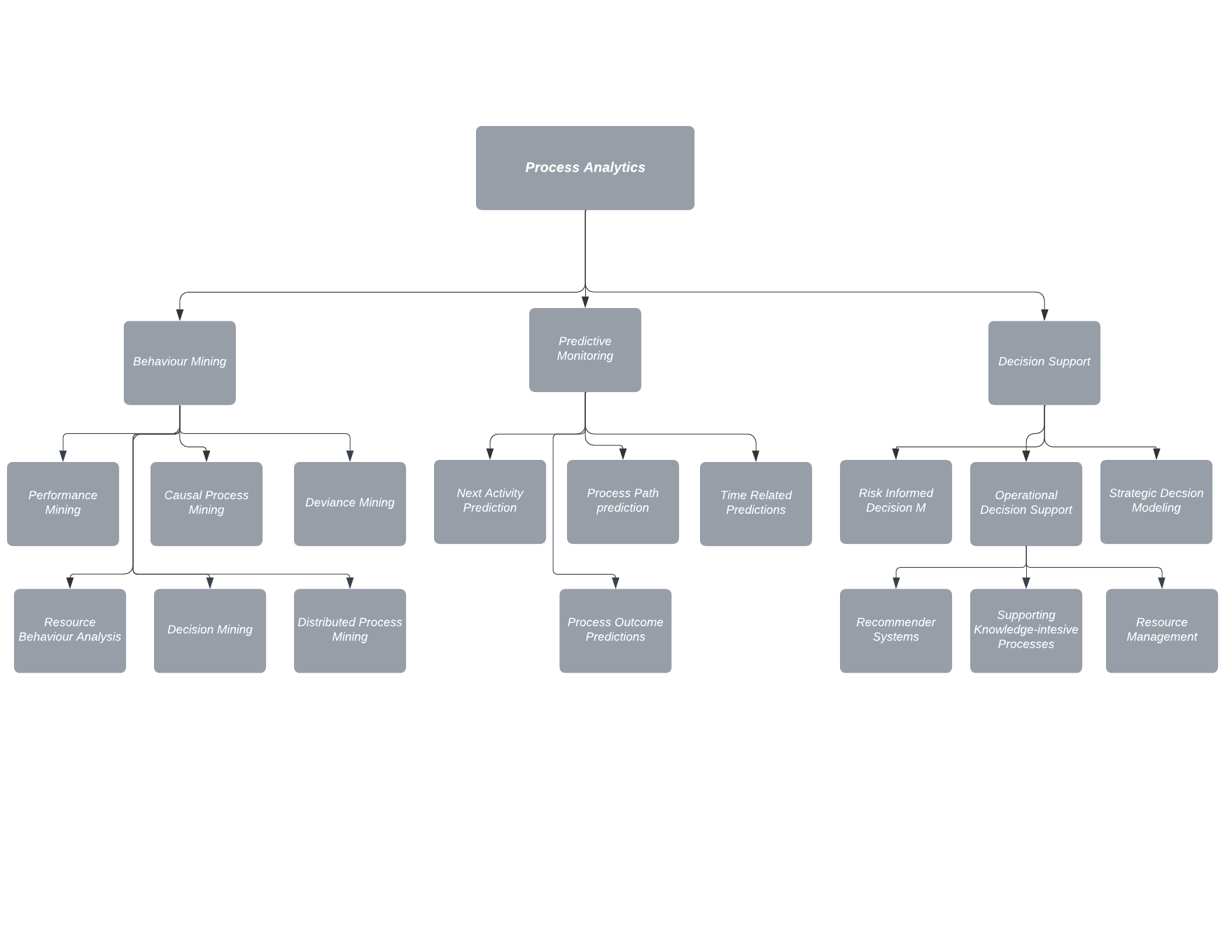}
    \caption{Proposed Taxonomy}
\end{figure}

\section{Mining Process Behaviour} 

In a given enterprise, business processes, when executed often leave an operational data footprint in the form of logs, documentation, and various data artifacts containing insights and knowledge that are of considerable business value. Process mining techniques can leverage this execution data to mine actionable knowledge, discover insights about performance bottlenecks, frequent defects, their root causes and other sources of inefficiencies. This allows organizations to gain behaviour visibility in order to optimize  process performance and support both operational and managerial decision-making \cite{benatallah2016process} \cite{dumas2018business}. Enterprises adopt process mining tools that typically support business process improvement by offering techniques for process automation, auditing and compliance checking, and recently digital transformation  \cite{kerremans2018market}.

Traditional process mining techniques cover various process-related perspectives. e.g control-flow perspective, organizational perspective, case perspective and time perspective. In behaviour analysis, we start by understanding the current ‘as is’ state of processes. Here process discovery techniques can for example by leveraged event data to for example reveal (in terms of an abstract process model) the common workflows used by an enterprise for executing various types of functionality. i.e. we describe how process activities currently are being performed by mining the process model from event log.  The control-flow model can then be extended with other perspectives to obtain a holistic view covering all process-related aspects and gain a complete understanding of current process behaviour. The insights and knowledge gained during this process can be used for the redesigning processes and offers a means for improving process performance such that it leads to positive (value-adding) process outcomes \cite{zur2015business}.

Behaviour analysis also aims to provide an analytical support layer that addresses the information needs of process field analysts \cite{klinkmuller2019mining}. Such support involves combining multiple analytic techniques to: \textit{(i)} become familiar with the dataset at hand in order to answer a particular question \textit{(ii)} obtain performance statistical summaries(e.g wait times, historical cycle times) using aggregation, correlation and evaluation techniques \textit{(iii)} uncovering insights, patterns and discovery of hidden relationships among various organizational elements/artifacts \cite{hamilton2015field}. During the 'exploratory data analysis' phase, process analysts often engage in techniques like deductive (hypothesis-based) and inductive (pattern-based) reasoning to formulate and to answer one or more questions \cite{hamilton2015field}.  Traditional process mining and discovery algorithms \cite{klinkmuller2019mining} for generating procedural models (like Petri nets,  causal nets, BPMN models process trees) represent the control-flow of the process and can give a big picture view of existing deployed business process models. Visualizing the control-flow or process models through process discovery is a good starting point for process analysts to perform further analysis by asking questions in an iterative manner \cite{van2011process}. The initial analysis then guides the process mining activities as information needs emerge during familiarization and discovery phase \cite{van2011process}\cite{klinkmuller2019mining}.  In \cite{klinkmuller2019mining}  Muller et al. have investigated challenges faced by process analysts in practice. According to their findings, analysts spend significant time investigating the case perspective (which focuses on properties of cases) and organizational perspective(which focuses on organizational resources).  In \cite{van2009process} Van der Aalst has also shared some lessons learned after conducting practical real world process mining projects.

Process mining algorithm's performance is traditionally measured by how well it achieves pareto-optimality of the mined model in terms of various properties such as fitness and precision with respect to the available event log. The algorithm has an additional goal of achieving generalization on future process instances. In many practical settings, the target search space of models is quite large for an exhaustive search; therefore, process mining algorithms, enforce a specific representational bias to to make a trade-off (e.g. between higher fitness and lower precision).  In many real-world settings, process behaviour is not completely captured or available for mining in the event logs  \cite{augusto2018automated} as these logs are often \textit{noisy} and \textit{incomplete}. Process discovery algorithms when applied to real-world complex event logs often produce either noisy or incomprehensible models that either poorly fit the event log (low fitness) or over-generalize it (low precision or low generalization) \cite{augusto2018automated}.

Process mining techniques discussed in this section along with classic methods like conformance checking and process enhancement, collectively help us construct an enriched process model through the offline analysis of all available process-related data. This enriched process model provides us with a holistic process view, covering all relevant aspects of the process at hand and is helpful for: \textit{(a)} supporting managerial decision-making and generating insights about performance improvement \textit{(b)} constructing a simulation model which captures the on-ground reality And; \textit{(c)} Performing ''what-if'' analysis using the simulated model. \textit{e.g., if I allocate a given resource set to the next task, what would the predicted completion time for the process instance be?}. Next, we will focus on reviewing process mining techniques aimed at behaviour analysis of all the available structured and unstructured process-related data, in order to: 

\begin{itemize}
    \item Assist business analysts in gaining visibility and understanding, of process behaviour captured in process execution logs and other process-related data by Answer a wide range of process-related questions.
    \item Uncover patterns, extract insights and discovering hidden relationships among various organizational elements/artifacts.
    \item Support managerial decision-making by mining actionable knowledge and insights related to performance bottle-necks, optimal resource allocation and uncovering root-causes that lead to undesired process outcomes. 
\end{itemize}

\subsection{Process Performance Mining}
 
Prior to process design, organizations often engage in an exercise of strategic planning by using balanced scorecards where the mission and strategy of an organization \cite{goodspeed2004translating} are translated into a set of performance measures (also known as KPIs). Such performance measures, can help organization keep track of progress towards the specific defined organizational goals. Moreover, for organizations implementing business process management, organizational objectives are used to formulate process performance measures, characterized by metrics like per-instance cost, cycle time efficiency (CTE), resource utilization, quality of service and so on \cite{zur2015business}. Several performance measurement systems exist, and appropriate ones can be picked depending on the organizational objectives or strategic success factors \cite{vom2010handbook}. Devil’s quadrangle can for example, measures dimensions such as process cost, quality (e.g. visiting frequencies, error rates) and cycle time (also known as throughput time)\cite{jansen2007performance}. Jansen-Vullers et. al \cite{jansen2007performance} have proposed a framework for quantifying the impact of best practices and discuss different dimensions for performance measurement. Several domain-specific performance reference models also exist. e.g Supply Chain Operations reference model, IT infrastructure Library (ITIL) etc.
 
Performance analysis techniques allow us to extract, analyze and enhance existing process models by identifying performance improvement areas and mining KPI values using the available process logs. Early identification of large deviations between the planned and actual KPI values allows organisations to take corrective actions to achieve the desired goals.  Performance analysis techniques include Bottleneck Analysis (where we identify activity, resource and waiting bottlenecks), workload and demand analysis (where we analyze resource usage to identify under-utilized or over-utilized resources), rework analysis (where we identify errors or defects) and over-processing analysis. Organizations can leverage these techniques to identify a range of issues related time, costs and quality based on event log data. These insights can later be used in the \textit{process redesign} phase for improving or enhancing the apriori process model.  We note that process performance analysis has several titles in the literature. e.g Business Process perspective,  Performance perspective, Process Performance Management etc \cite{hornix2007performance} \cite{zur2015business}.

Performance analysis techniques can be divided into \emph{Qualitative or Quantitative analysis} \cite{dumas2013business}.  In Qualitative Process Analysis, we can perform Value-Added Analysis(which aims at identifying waste) and Root Cause Analysis(e.g cause-effect analysis and why–why analysis).  Here we measure process performance using metrics like per-instance cost, cycle time efficiency (CTE), resource utilization, quality of service, and so on \cite{zur2015business}.  Quantitative process analysis is based on historical process execution data or simulation models and involves using techniques like flow analysis, queuing theory and process simulation \cite{dumas2013business}. \emph{Process simulation} is a popular technique for quantitative analysis of process models where synthetic data based on hypothetical instances is used to estimate cycle times, average waiting times and average resource utilization. Similarly, \emph{flow analysis} provides a set of techniques that utilize activity performance data to estimate the process's overall performance but ignores the resources utilization aspect from the analysis \cite{dumas2013business}. 

\emph{Business process monitoring} deals with event analysis and support of real-time executions of a given process. It aims to understand the present process performance by presenting a real-time picture of key performance indicators(often using dashboards) such as mean execution time, resource utilization or error rate etc.\cite{dumas2013fundamentals}. Process monitoring tools help determine how well business processes under consideration are aligned with the goals of an organizations. They enable continuous, real-time monitoring of processes based on performance-related KPI values of a given process instance to track various business objectives.  Process monitoring tools like performance dashboards can monitor the actual KPI values(e.g the distribution of cycles times, visualize the activities and execution patterns, and total duration) of a business process\cite{tardio2015obtaining}. Several software vendors offer various performance analysis features where we can find the causes of these deviations(when they happen) and project performance analysis results onto process models to show bottlenecks, service performance levels, throughput times, and frequencies \cite{hornix2007performance}. 

\subsection{Causal Process Mining}

%Lastly, discovering cause-effect relationships in event data is useful for root cause analysis of process performance issues but remains a challenging problem. 

To inform decision making under uncertainty, Businesses often draw on previous experience and recorded data to understand the potential downstream effects of certain specific decisions and actions. Historically, to determine the effect of a given intervention on a particular process would require either guesswork or launching numerous A/B tests, which can be costly and time consuming \cite{kohavi2017online}.  Theory of causality provides a better alternative for potential downstream effects of decisions that are being considered \cite{pearl2009causal}. Causal process mining seeks to use the process execution logs to discover and quantify cause-effect relations. Causal Process Mining can answer the fundamental question: \textit{What changes, if implemented, will cause an improvement to the process?}. Existing process discovery techniques allow us to discover correlation but not causation. In causal analysis we try to develop an understanding that goes beyond the control-flow perspective.  We are interested in understanding the outcomes(desired or undesired effects) associated with particular action(s) (interventions)  taken during the course of process execution. There has been much recent interest in studying Causal analysis and Causal reasoning in the data science and machine learning research communities. 

Leveraging process data would allows us to infer future process state based on what happened in the past and use that evidence to establish causality. However, pinning down causal effects rigorously is challenging. We briefly cover some of the recent contributions to process analytics in this section. Koorn et. al. \cite{koorn2020looking} have proposed a method that uses statistical tests to discover action-response-effect patterns. Their method can identify causal relations between responses (interventions) and effects for certain pre-defined sub-populations. This can support the decision making processes by giving insights into how certain actions lead to desired outcomes (e.g improved performance). In \cite{Qafari2020counterfactual} Qafari et. al. showed that root cause analysis using structural equation modeling is useful for testing if a predetermined causal relation (identified earlier by a process analyst) holds.  Bozorgi et. al. \cite{dasht2020process} proposed an approach that extracts recommendation rules from event logs. Causal effect can then be assessed and rules with highest incremental effect (uplift) can be used as recommendations in the form of interventions that can influence process outcomes positively. Their framework also computes a cost-benefit model of a particular intervention, identifying particular cases to which it is applicable (case-level recommendations). Agarwal et. al \cite{narendra2019counterfactual} take a similar approach for structural causal model discovery and performing counterfactual reasoning. Discovering cause-effect relationships in event data is useful for root cause analayis of process performance issues but remains a challenging open problem. Hompes et. al. \cite{hompes2017discovering} propose a method which uses Granger causality to generate a graph of causal factors explaining factors (or their combinations) that may affect process performance. 

Causal Analysis techniques has also been used to understand process deviation, explain predictions and make recommendations. We will discuss these techniques in next sections. 

\subsection{Deviance Mining}

Business process variant (also known as deviance analysis or drift detection) refers to process instances that maybe deviate from the desired course of execution, resulting in an unexpected or unplanned outcome. Variations may occur due to contextual/environmental factors, human factors or because of explicit decisions made by process participants. In a given process log, process variants are a subset of instances that violate the behavior prescribed by the model and can distinguished based on a certain characteristics they are correlated with. e.g. representing violation compliance rules or missing set performance targets. In deviance mining, we are interested in identifying various variants of the processes that may exist and diagnosing the root-causes of process variations by analyzing or comparing two or more event logs. i.e \textit{Given a set of event logs of two or more process variants, how can we identify and explain the differences among these variants?} \cite{taymouri2019business}.  Understanding the factors leading to variation can help managers make better decisions and improve overall process performance.

Several methods have been proposed in the last decade to analyze the execution logs to identify and explain differences between two or more process variants. One class of techniques uses \textit{frequent pattern mining} which typically takes as input two event logs (corresponding to two variants of a process) and produces as output a list of differences (with respect to established performance objectives). Machine Learning techniques on the other hand can classify instances as \textit{``normal''} or \textit{``deviant'' } when trained with labelled examples. In \cite{nguyen2016business} Nguyen et al. have surveyed sequence classification and mining techniques. They divide the techniques into those based on frequent pattern mining and rest based on discriminative pattern mining. They also provide benchmark comparisons of representative techniques using various real-life event logs. Similarly, Taymouri et al. \cite{taymouri2019business} have tried to present a unified view of variant analysis techniques by surveying and classifying existing methods based on data, type of algorithms and analysis performed. Several methods rely on identifying frequent patterns, while others use generative approaches to discover and compare models of process variants. They have categorized process variant analysis outcomes as Rule-based, model based or descriptive (representing discrepancies and behavior of the different process variants.  Modeling and management of process variants is another important challenge that contemporary BPM tools do not adequately support \cite{reichert2015lifecycle}. Existing tools require that variants be specified in separate process models or 'expressed in terms of conditional branches within the same process model'.

%Such tools can also Assist in detecting the root causes of possible deviations in the values of the monitored KPIs

\subsection{Decision Mining}

%A business process consists of several decision points and Sub-optimal decisions, for example, picking the wrong execution paths can lead to cost over-runs and missed deadlines. T

Operational Decision-making associated with frequently executed processes and cases can significantly impact process outcomes. Traditionally, a business process model specifies activities within which the decision-making occurs.  Decision logic can define the specific logic used to make individual decisions, such as business rules or executable analytic models. Decision modelling can represent complex, multi-criteria business rules and thus provides another perspective, creating a bridge between business process models and decision logic models \cite{figl2018we}. 

\textit{Decision Model and Notation (DMN)} is a modeling notation for decisions, published by \textit{OMG} and allows for decoupling between decisions and control flow logic \cite{model2016notation}. It provides an understandable symbolic representation of operational decisions and supports supports decision management, and business rules specification \cite{kluza2019understanding}.

% Similarly, a business rule in an organization can be represented by a decision rule.Decision rules can also be seen as functions that take as input a number of parameters representing decision criteria and compute the best path \cite{catalkaya2013enriching}. 

Decision mining aims at mining the Decision model along with along with associated decision logic. The Decision model will define the decisions to be made in tasks (defined by the process model), their interrelationships, and their requirements for decision logic. Process execution data, depicting underlying rules (governing the choices) can be mined for frequently made decisions within an organization. However, decision mining is challenging due to several factors (e.g incompleteness of available data). Decisions are often dependent on personal expertise and contextual factors, information that may not be present in event logs.  Decision mining techniques can identify and derive decision rules from relationships between process context, path decisions, and process outcomes. Several techniques proposed in the literature has tried to tackle the problem of decision mining partially.  In mannhardt et al. \cite{mannhardt2016decision} propose a decision-tree based learning method to discover overlapping decision rules from event data. Similar to decision mining, Batoulis et. al. \cite{batoulis2015extracting} have proposed a method for extracting decision logic from process models. The paper explains a semi-automatic approach where execution logs are not required and decision points can be replaced by generating a dedicated decision model, allowing decision logic to modeled separately from process logic. Rozinat et. al. \cite{rozinat2006decision} propose a method for decision point analysis that determines how data attribute values(or data dependencies) affect the routing of a case. Their technique can identify decision points in Petri net models and has been implemented as a ProM Plug-in. Leoni et. al \cite{de2013data} extended the technique and proposed a general-purpose method for discovering \textit{Branching Conditions} (where atoms are equalities or inequalities consisting of multiple variables and arithmetic operators). Similarly, Bazhenova et. al. \cite{bazhenova2016discovering} use decision tree classification to derive of DMN based decision models. Their techniques are able to extract not only control flow decisions but data decisions and dependencies. Overall, a comprehensive set of methods and frameworks is required that assist orgnaizations with dynamic management of decisions via analyzing, modelling and improvement efforts.   
Such methods should be able to leverage available process data and mine Decision model along with along with decision logic and dependencies between decisions and data elements .

%This helps organizations in modelling/analyzing decisions that affect the control-flow and routing of a process instance \cite{rozinat2006decision}. 

%The problem of finding the branching conditions and understanding the data-flow has received less attention. i.e. a lot of focus of process mining has been on finding the control-flow relations between events or task and less attention has been paid to the data-flow perspective. 

\subsection{Resource Behaviour Analysis}

Human resources often take the role of knowledge workers and play a major role in modern organizations where  processes are increasingly complex have knowledge-intensive nature \cite{di2015knowledge}. In such knowledge-intensive scenarios, Resources (especially human resources) play a critical role in deciding the overall process outcomes.  Event logs contain rich information about the task, resource and process outcome and can be leveraged to optimally realize the process goals \cite{Renuka2018}. Resource analysis (also known as organizational or resource mining/perspective) aims at analyzing event logs for extracting insights about organizational resources, evaluate resource performance and understand past resource allocation decisions in order to find areas of improvement \cite{pika2017mining}. Over the past decade, various resource analysis techniques have been proposed to tackle problems like organizational structure discovery, classification of users in roles, discovery of organizational models, social network analysis (SNA), resource allocation, analysis of information flows between organizational entities and role mining \cite{zhao2014process} \cite{burattin2013business} \cite{song2008towards}\cite{Renuka2018}. 

Optimal allocation of resources to process tasks can significantly impact the overall process performance. Comparing and tracking the productivity of human resources is a challenging problem due to biases and lack of objectivity. Sindhgatta et al. \cite{sindhgatta2015learning} propose a method that uses process execution logs to identify resource allocations decisions which result in good outcomes(measured in terms of quality of service).  In a similar work, Sindhgatta et al. \cite{sindhgatta2014analysis} investigate the variation in resource efficiencies with varying case attributes and show that process outcomes are dependent on various contextual factors like complexity of work, task priority and capabilities(expertise) of the resources involved. 

Accurately measuring resource behaviour allows us to effectively dispatching and suggest staffing policies that meet the contractual service levels (quality) of the service system and the business process. Huang et. al. \cite{huang2012resource} consider the problem of measuring resource behaviour from different perspectives such as preference, availability, competence and cooperation. Similarly, several methods for extracting useful knowledge about resource performance from event logs have been proposed in the literature. Pika et al. \cite{pika2017mining} propose a technique based on data envelopment analysis for analysing resource productivity using event logs. This technique is often used to measure the efficiency of companies. Linh ly et. al. \cite{ly2005mining} study the problem of mining staff assignment rules from event-based data using an organisational model. Similarly, Senderovich et. al. \cite{senderovich2014mining} explore mining of resource scheduling protocols from recorded event data. Related to the problem of resource assignment and allocation Cabanillas et. al. \cite{cabanillas2013priority} have studied the challenge of resource ranking and prioritization.

The success of process outcomes also depends on resources interaction.  Social collaboration patterns between human resources can significantly impact process outcomes as optimal cooperative resource behaviour leads to enhanced service quality and process performance. Improving process performance by analyzing relationships and mining collaboration patterns between organizational entities is one of the key goals of resource analysis. It is based on the premise that process output not only depends on capability but compatibility as well. Social network analysis \cite{van2004mining} is one approach for understanding organizational structure and analyze relationships between originators involved in processes. Such an analysis is useful for many reasons such as improving handover relations, improving resource compatibility etc. In \cite{kumar2013optimal} Kumar et al.  propose a modeling technique that captures the compatibility between resources at the time of task assignment.  Schonig et. al. \cite{schonig2015mining} proposed an approach to extract declarative process models. Their technique allows modelling of organizational relations using rule templates that can be represented in textual Declarative Process Intermediate Language (DPIL).
 
 %The goal is to either structure the organization by classifying people in terms ofroles and organizational units or to show the social network
 
RBAC (Role based access control) is used to manage resources in workflow area and is a  useful model for managing resources\cite{kuhlmann2003role}. Event logs can be used to to mine role based access control (RBAC) models,  identifying the privileges or authorization of resources \cite{baumgrass2011deriving} \cite{burattin2013business}. Roles refer to the assignment of  organizational agents to job functions within an organization. Discovering an optimal set of roles remains a useful problem to investigate. Vaidya et. al. \cite{vaidya2007role} define the perform as determining a role-based access control (RBAC) configuration and analyze its theoretical bounds. Similarly, frank et. al.  \cite{frank2013role} propose a probabilistic solution to learn the RBAC configuration and show how it generalizes well to new system users for a diverse range of data. 

%\subsection{GOal Model mining}
%\subsection{Strategy mining}
%subsection{Enterprise Architecture Mining}
%subsection{Context Mining, Mining Process Provisioning insights}

\subsection{Distributed Privacy-Preserving Mining}

Modern organizations routinely deploy process analytics, including process discovery techniques on their process data, both to gain insight into the reality of their operational processes and also to identify process improvement opportunities \cite{augusto2018automated}. Process analytics techniques such as process discovery play an important role in mining event data and providing organizations with insights about the behaviour of their deployed processes. However, in many practical settings, process log data is often geographically dispersed, may contain information that may be deemed sensitive and may be subject to compliance obligations that prevent this data from being transmitted to sites distinct to the site where the data was generated. Traditional process mining techniques operate by assuming that all relevant available process data is available in a single repository. However, anonymising, giving control access and safely transferring sensitive data across organization/site boundaries while preserving priacy guarantees is non-trivial. However, in many practical settings, process log data is geographically dispersed and can contain information that may be deemed sensitive. A classical example of a privacy-preserving process mining problem of the first type is from the field of medical research involving impediments to data migration. Consider the case that a number of different hospitals wish to jointly  mine  their  process  logs  for  the  purpose  of  medical research, but are faced with regulatory and legislative compliance hurdles that prevent clinical process histories being shared
across health jurisdictions (hospitals, health districts, national boundaries etc.). Hospitals are therefore restricted from ever pooling their data or revealing it to each other leading to small dataset available for knowledge extraction.  This negatively impacts the confidence with which clinicians might deploy the results thus obtained. Our inability to migrate clinical process data also implies that we miss out on the opportunities for extracting higher-impact insights that might have been possible if data from multiple health jurisdictions could have been analysed in juxtaposition \cite{lenz2007support}.

Traditional process mining techniques operate by assuming that all relevant available process data has been curated into a central site for analysis. However, anonymising, giving control access and safely transferring sensitive data across organizations is non-trivial. Moreover, organizations face legal constraints, risk of data breaches (or hacks) along with data integration challenges, preventing them from building a centralised data warehouse \cite{dunkl2011assessing}. This leads to a scenario where event-log data is present in organizational silos and distributed among several custodians, none of whom are allowed to share/transfer their sensitive data directly with each other \cite{lang2008process}. Mining process data in such cross-silo settings can prove to be invaluable for providing relevant operational support to organizations if privacy guarantees can be offered \cite{jensen2012mining}.

\textit{Differential Privacy} provides us with a formal privacy notion for datasets that are released publicly or might come in contact with potentially malicious adversaries \cite{mcsherry2007mechanism}.  It is considered as the \emph{de facto} standard for ensuring privacy in a variety of domains.  The definition proposed by Dwork et al. \cite{dwork2014algorithmic} offers a mathematically rigorous gold standard for ensuring privacy protection when analyzing datasets like process logs(or results of a randomized algorithm) that might contain sensitive or private information. We modify the definition slightly for event logs: 

\textbf{Definition 1: Differential Privacy (adapted from \cite{dwork2014algorithmic})} A randomized mechanism $\mathcal{M}: \mathcal{D} \rightarrow \mathcal{R}$ with a domain $\mathcal{D}(e . g .,$, possible event logs) and range $\mathcal{R}(e . g .$, all possible trained models $)$ satisfies $(\epsilon, \delta)-$ differential privacy if for any two adjacent process logs $l, l^{\prime} \in \mathcal{D}$ and for any subset of outputs $S \subseteq \mathcal{R}$ it holds that $\operatorname{Pr}[\mathcal{M}(d) \in S] \leq e^{\epsilon} \operatorname{Pr}\left[\mathcal{M}\left(d^{\prime}\right) \in S\right]+\delta$

Two process log $l$ and $l^{\prime}$ are defined to be adjacent if $l^{\prime}$ can be constructed by adding or removing a single instance(entry) from the log $l$. By bounding the potential worst-case information loss, the above definition provides us with a strong formal privacy guarantee.  Formally, under the $(\varepsilon, \delta)$-differential privacy definition, we measure Differential Privacy properties of our method by epsilon and delta values.  Epsilon($\epsilon$) is the privacy loss parameter in differential privacy and is inversely proportional to the amount of noise added. i.e Lower values of $\varepsilon$ imply stronger privacy guarantees. A Differentially private mechanism typically involves using a randomized mechanim that perturbs the input dataset, intermediate calculations, or the outputs of a function, using a calculated quantity of noise (usually at the cost of utility) \cite{dwork2006our}. Such a mechanism is considered private if it hides the isolated contribution of any single individual in the databases. i.e removing a single entry will not result in much difference in the output distribution \cite{acs2012dream} \cite{dwork2014algorithmic}. 

%arties, collaborating to compute a common function of interest, without revealing their private inputs to other parties \cite{goldreich2019play}. The protocol is considered secure if, at the end of the computation, parties learn nothing but the final result and no other information. Secure Aggregation is a class of Secure Multi-Party Computation algorithms wherein a group of mutually distrustful parties $u \in \mathcal{U}$ each hold a private value $x_{u}$ and collaborate to compute the aggregate value(such as  sum $\sum_{u \in \mathcal{U}} x_{u}$) without revealing to one another any information about their private values except what is learnable from the aggregate value itself \cite{bonawitz2016practical}\cite{clifton2002tools}. Secure aggregation in a federated learning setting presents its own unique set of challenges, which several recent works have tried to tackle \cite{bonawitz2017practical} \cite{so2021turbo}. 

In Privacy-Preserving Distributed Process Discovery, our goal is to discover a global process model by privately mining multiple distributed process log independently and share only the resulting insights from each analysis. i.e mining a differentially private process model, without ever pooling the data to a central site, in a way that reveals nothing but the final discovery process model to the participating organizations. 

Most  typical  methods  presented in the literature rely on some form of data transformation in order preserve user  privacy.  These  techniques  are  a  trade-off  between  information  loss  and privacy\cite{bamiah2012study}. 
Techniques based on Secure Sum Protocol \cite{clifton2002tools} permits a network of nodes to transmit a numeric sum (from node to node) to
which each node adds a node-specific number without having any node being able to compute what the
individual contributions of the participating nodes were. The final node obtains the sum of the numbers
contributed by all of the participating nodes, again without being able to compute what the individual nodespecific numbers were. This protocol can be used to create distributed versions of most process mining algorithms.  Elkoumy et al. \cite{Elkoumy2020} propose an architecture based on Sharemind, which uses multiparty compute to mine a directly follows graph. However they don't provide differential privacy guarantees.

\subsection{Knowledge-Centric Process Mining:} 

 In process mining, while a lot of emphasis has been on analyzing and extracting process insights from the observed behaviour logged in event logs, the knowledge dimension associated with business processes has received very little attention \cite{di2015knowledge}. i.e Current process mining techniques are self-contained and have minimal capacity to leverage and reasoning using prior knowledge. In practice, this means process mining algorithms can't perform inference that goes beyond the implicit knowledge which is recorded in the event logs. Traditional mining techniques focus on mining behaviour that inherently cannot represent all of the cascading hierarchical structure representing complex real-world processes \cite{van2009process}. These algorithms can't reason about abstract relationships between various objects involved in the process. For example, easily-drawn inferences that people can readily answer without direct training like \textit{smoke is seen so there must be fire happening} cannot be inferred by the current process mining methods. This leads to an incomplete understanding of process behaviour where process analysts are left trying to abstract, simplify, and even leave out key relationships needed for complete understanding of process behaviour.

Process knowledge has many faces and it will differ from other forms of organizational knowledge as it will be highly contextual, sometimes tacit and relevant to a particular domain.  Organizations have realised that knowledge and processes are interlinked and should explicitly be made a key component of business processes \cite{barclay1997knowledge}. Knowledge is more complex than simple data or information with an additional characteristic of being subjective, as its often tacit in nature (e.g obtained with years of experience).  Process knowledge has many faces and it will differ from other forms of organizational knowledge as it will be highly contextual, sometimes tacit and relevant to a particular domain.  Sometimes this knowledge is documented but even then it exists in silos of unstructured documents, often halting the progress of processes to consult any knowledge bases that might exist. 

For organizations, having access to the right knowledge at the right time is crucial to effective decision making as it allows companies to solve problems quickly, diffuse best practices amongst employees, cross fertilize ideas and enable them to stay competitive \cite{van2002model}. Managing Process knowledge requires a deliberate and systematic approach which should be reflected at all levels of organization. Organizations employ various knowledge managements tools to capture and create the knowledge that is either explicit or tacit by sometimes interviewing experts, telling war stores, and apprenticeship style training programs. Despite several attempts, modern organizations find capturing managing the available knowledge(in terms of capturing, codifying, and sharing) difficult and rely on knowledge workers for know-how. This leads to organizations routinely forgetting the lessons learned during the past when knowledge workers leave or switch projects. Overtime organizations have realised the importance of deliberate and systematic approach for creating a culture where company’s knowledge base in maintained \cite{von2014complete} and  many expert systems, and Enterprise knowledge management systems have been proposed to capture knowledge effectively, categorize it and then make that knowledge available across an organization. Such tools can codify knowledge in the form of wikis, cognitive maps, decision trees, knowledge graphs etc. These movements were largely unsuccessful\cite{easterby2011handbook}. 

Knowledge management has largely been ignored in recent years. Solutions are needed that tackle knowledge management challenges and help organization extract actionable knowledge from all available process related data(e.g extracting meaningful knowledge from unstructured documents) and make it widely available inside the organisation. Similarly, \textit{Common-sense reasoning} has been highlighted as one of the major challenges for the process analytics research community \cite{calvanese2021process}. It follows a broader trend in AI research where the need for solving complex tasks by  incorporating knowledge and \textit{common-sense reasoning} has been repeatedly highlighted \cite{davis2015commonsense}. 
Furthermore incorporating domain knowledge by means of constraints can support process analysts in their efforts to fully understand executed process behaviour recorded in real-world event logs and  improve the outcome of process mining activities.

\subsection{Discussion:}  

The topics covered in this section, along with classic topics like \textit{process discovery} continue to provide opportunities for research contribution. We note that even with the availability of various state of the art process discovery methods, understanding business processes and resource behaviour from process logs alone remains a challenging problem\cite{benatallah2016process}. Furthermore systematic adoption of process mining in challenging domains like healthcare, also remains a challenge \cite{munoz2022process}. Process mining techniques can be of great value to answer process-related questions and is often a starting point for process analysts, however, more support is needed to address the challenges faced by process analysts. Carmona \cite{carmona2020process} has highlighted some of the open problems and research challenges associated with process discovery and conformance. Their findings show that the existing mining methods struggle to deal with challenges like spaghetti models, concept drift and identifying events that occurred at different levels of abstraction.  We also observe that traditional process mining and analysis techniques have not given significant attention to the analysis of resource behaviour and its affects on process outcomes. Resource analysis deserves more attention from the research community as knowledge about resource behaviour can be used for effective planning, strategizing and gaining insights which lead to better overall process performance. In \cite{vom2021five} Brocke et al. present an enterprise framework for analyzing the effects of process mining that emerge at various levels of an organization. Lastly, by understanding contextual factors and gaining detailed insights about how processes are being executed within a particular context also remains an interesting challenge.

%We briefly discuss relevant areas (in terms of research value) that future studies should focus their investigation on:  

\textit{Data Challenges:} Process mining techniques are primarily reliant on process logs, which don't always explicitly capture all the behaviour of past executed processes\cite{augusto2018automated}. Such logs are susceptible to domain gaps, data bias (due to incompleteness) and quality issues (due to noise and erroneous data recordings). Many real world processes are unstructured in nature and for these processes most state-of-the-art process discovery algorithms produce, hard-to-interpret, spaghetti-like models which poorly fit the event log. This negatively influences the usefulness of the discovered process model. Often times discovered models are hard to interpret from a process analyst perspective while also prone to under-fitting or over-fitting the given event logs, offering only minuscule support for improving process outcomes. Furthermore, a particular challenge in process mining is the management of business-process variants and contemporary business process management tools do not provide adequate support for modeling and management of process variants \cite{taymouri2021business}. For complex domains like healthcare, where improving clinical outcomes can directly impact the quality of life for patients, this implies that process analysts miss out on the opportunities for a complete understanding of the underlying process behaviour and subsequently extracting higher-impact insights. 

Event data can come from various heterogeneous data sources and in several data formats.  These logs are rarely in the desired format and form. It is common for process analysts to apply a series for pre-processing steps for consolidation, verification(checking for errors and inconsistencies) and transformation. To identify an event trace representative of a process instance, is challenging and is often done manually using techniques like extraction, correlation, and abstraction of the event data by the process analyst \cite{diba2020extraction}. There is also the challenge of \textit{'Big data'}, where data to be analyzed is of huge volume, velocity and variety, making it unfeasible to be analysed with traditional analysis tools. Other challenges include the diversity of formats and nonstandard data models \cite{sakr2018business}. To maximize the utility of available data, it is sometimes useful to build a data lake that provides a single consolidated view of organizations' datasets. Data lake makes the preparation and analysis(by querying) of data more accessible. Such data can also include relevant context in which business processes were executed (this could be the context for an entire process or an individual task) and enable organizations to gain insights that help improve existing processes.

As observed in various other fields of AI and machine learning, an important driver of measuring progress is precisely defining Process Analytic tasks(using mathematically rigorous definitions) and building useful benchmark datasets that can be used for performance comparison \cite{blagec2020critical}.  The progress of the field depends on the cycle of identifying real-world problems, researchers proposing novel techniques and industry adaptation of such techniques. It is therefore important to make quantifiable progress by benchmarking process analytics tasks on standard datasets with well-defined performance metrics. We observe that many process mining published papers often lack strong reproducible experiments, resulting in poor comparability and relative merits of the proposed approach. So far, only in predictive process analytics, we have seen the adoption of rigorous evaluation criteria. We refer to work by  \cite{augusto2018automated} as an example of such an effort where Augusto et al. review various state-of the art process discovery methods and benchmark the performance of various automated discovery algorithms. We hope future studies will follow the same approach where they critique and compare proposed methods against existing state-of-art process mining techniques via standard benchmarks.

\section{Predictive Process Monitoring}

Predictive business process monitoring is a family of techniques concerned with predicting the future state, outcomes and behaviour of ongoing cases of a business process\cite{teinemaa2019outcome}. For organizations, process monitoring is a useful capability that enables operational support for near real-time monitoring of processes and allows them to take preventive measures that help avoid conformance violations, undesired deviations and prevent delays.

Predictive analytics can be also be viewed as, computing a set of functions or a set of computer programs that carry out computation, over a (partially executed) process instance to perform continuous monitoring of process instances. Such methods can leverage historical(completed) executions logs, such as event logs, along with available contextual data, in order to: recommending appropriate actions at each stage, early detection of process variation or anomalies(for fraudulent behaviour), predict various properties of a case instance, help perform early risk assessments  and assist with resource allocation decision\cite{weinzierl2020empirical}. 

The research goals of the field of Predictive Analytics can be framed as following research questions: 

Given an an event log of a completed business process execution cases and the final outcome for each case, Can we: 

\begin{itemize}
  \item Predict the next best activity to execute
   (in order to achieve optimal outcomes)? or Predict the outcome of a single activity
  \item Predict the entire sequence of activities leading to the process end? 
  \item Predict if the running process instance will meet its performance targets 
  \item Predict the total remaining time to completion for a process instance?
  \item Predict the performance outcome of an incomplete case, based on the given (partial) trace   (redundant)
  \item Predict the cycle time of a given process instance ?
  \item Estimate the likely cost that will be incurred in executing the remainder of the process? 
\end{itemize}

Machine learning allows us to build models from past experiences embedded in the process data and use it to provide real-time or near real-time decision support. Over the years many machine learning techniques, such as regressions, support vector machines etc. have been used to build predictive models from historical data. In Predictive Process Monitoring,  we study methods for building predictive models that utilize historical execution traces to predict the likelihood of events and forecast outcomes such as \emph{next activity} prediction,  \emph{process outcome} prediction or \emph{remaining time} prediction \cite{maggi2014predictive}. Predictive analytic methods can determine factors that influence the process outcomes and violate performance targets \cite{wetzstein2009monitoring}.

\textbf{\textit{Deep Learning:}} Recent advances in neural network architectures and availability of large datasets has led to the popularization of using \emph{'deep learning`} methods for predictive analytics. Deep Learning methods have been proposed for predicting how the future of a given process instance will unfold and the likely occurrence of certain events. \emph{Deep Learning} methods are particularly good at discovering intricate structure and robust representations from large quantities of raw data, thus significantly reducing the need to handcraft features which is typically required in traditional machine learning techniques \cite{lecun2015deep}. DNN's have a number of processing layers that can be used to learn representations of data with multiple tiers of abstraction. The different tiers are obtained by producing non-linear modules that are used to transform the representation at one tier into a representation at
a more abstract tier. \textit{Deep Convolutional Neural Nets} and \textit{Recurrent Neural Nets}
are two popular architectures of DNNs, that brought about breakthroughs in processing text, images, video, speech and audio \cite{lecun2015deep}.
\textit{Recurrent Neural Nets}, especially the \textit{Long Short-Term Memory (LSTM)} have brought about breakthroughs in solving complex sequence modelling tasks in various domains such video understanding, speech recognition and natural language processing \cite{lecun2015deep} \cite{schmidhuber2015deep}. LSTMs work by maintaining a dynamic short-term memory vector, which stores the summarization of historical events, from which the next activity can be predicted. It has been shown that LSTM can consistently outperform classical techniques for a number of process analytics
tasks such as predicting the next activity, time to the next activity and so  on.\cite{navarin2017lstm}.

Machine Learning methods require vector representations of input data and manually encoding or crafting features is a tedious task in practice. Deep learning methods have an advantage over such classical methods as they can
generalize well on various tasks without requiring explicit Feature engineering or
configuration tuning
\cite{arulkumaran2017brief}. Further these methods exhibit 'robustness to noise' and can show performance scaling as we input bigger and bigger datasets\cite{evermann2017predicting}. In Process analytics we can leverage the  deep learning methods to automatically find highly compact low-dimensional representations (features) of high-dimensional data. In \cite{Seeliger2021modeling} Seeliger propose recurrent neural networks (RNNs) based architecture that automatically learns vector representations of cases. They train the network to predict the contextual factors of the corresponding case. This allows us to incorporate contextual factors of a case into a single compact vector representation, later to be used for process mining.

Predictive process analytics techniques support managers in operational decision-making processes by for example taking remedial actions as business processes unfold.  Picking the best method depends on the domain, available datasets, choice of input features use train the models etc. For running process instances, making accurate and early predictions using limited computing resources still remains a challenge. Early accurate predictions of outcomes, explainability (reasoning behind why certain predictions were made) and prescribing actions to prevent undesired outcomes are some of the problems that are under active investigation by the research community. Various survey papers have tried to cover the literature on predictive analytics.  Marquez-Chamorro et al. \cite{marquez2017predictive} and Di Francescomarino et al.(2018) \cite{di2018predictive} classify the literature based on input data, classification algorithm and prediction target.  Similarly, \cite{teinemaa2019outcome} \cite{verenich2019survey} also survey the field by covering various datasets, propose task definitions and provide benchmark comparison of recently proposed algorithms. We briefly describe the relevant problems studied under predictive process monitoring: 

\subsection{Next Activity Prediction}

Next Activity Prediction refers to the problem of predicting the next trace suffix likely to occur during the execution. In Next Activity Prediction, we attempt to derive a process-agnostic machinery for learning to generate recommendations and predict the future behaviour of a given partially executed process instance while assuming minimal domain knowledge. Models are trained using historic data and after training, the input is using the observed prefixes of running process cases (event stream). The problem has been tackled using various machine learning techniques where it is often treated as a multi-class classification problem. The next event(of a process instance) can be represented as one class that the ML classifier can predict. 

Prediction techniques based on deep learning have a popular choice and have shown promising results\cite{evermann2017predicting} \cite{marquez2017predictive}. Such techniques are often motivated by the applications of deep learning to Natural Language Processing tasks (e.g langauge modeling).  Tama and Comuzi \cite{tama2019empirical} and Weinzierl et. al. \cite{weinzierl2020empirical} provide a comparison of architectures and encoding techniques for next activity prediction using real world logs. We can use several evaluation metrics(e.g Accuracy, F1-score etc.) to compare the methods. Brunk et al. \cite{brunk2020cause} consider the problem of context-sensitive process predictions(in business process monitoring) by employing evidence sensitivity analysis to determine if context is cause or effect of the next event during execution. This allow us to understand the impact that a context variable can have on a running instance and offers an explanation for understanding why a certain prediction was made. While LSTMs based techniques can theoretically deal with long event sequences, the long-term dependencies between distant events in a process get diffused into the memory vector. We therefore seek modeling methods that are more expressive and allow storing and retrieval of intermediate process states in a long-term memory. To tackle this, recently, Khan et al.\cite{khan2018memory} explore the application specific type of neural network known as the memory–augmented neural network (MANN) for the task of next event. In a typical setting, a MANN is a recurrent neural network (e.g., LSTM [8]) augmented with an external memory matrix. Differential Neural Computer architecture can be adapted the to account for a variety of tasks in predictive process analytics: \textit{(i)} separating the encoding phase and decoding phase, resulting dual controllers, one for each phase; \textit{(ii)} implementing a write-protected policy for the memory during the decoding phase.

%The proposed methods is evaluated using two unlabelled dataset for three predictive tasks: next-activity, time-to-event and suffix prediction and three labelled datasets to enable effective suffix recommendations.

\subsection{Process Path Prediction}

%Predictions Activity Sequence 

Predicting the evolution of running cases is an important problem and plays a key role in risk management, resource allocation and process improvement. Similar to Next Activity,  Process Path Prediction methods can predict possible paths of a running instance up to its completion\cite{verenich2016general}. Tax et. al.  \cite{tax2017predictive} and Evermann et al. \cite{evermann2016deep} explore the use of  Long Short-Term Memory (LSTM) to solve various process predictive problems.
Tax. et al.\cite{tax2017predictive} show the use of LSTM neural networks for predicting the sequence of the future activities. Similarly, Camargo et al. \cite{camargo2019learning} employ deep learning techniques to predict sequences of next events, their timestamp, and their associated resource pools.  
Considering context(as cause or effect) when making these predictions is also an interesting problem studied by Brunk et. al. \cite{brunk2020cause}. Tama and Comuzzi  provide a comprehensive empirical comparison and bechmarking for such techniques \cite{tama2019empirical}.

\subsection{Time-related Predictions}

Time-related Predictions problems such as Predicting the remaining cycle time, ompletion time and case duration, predicting delayed process executions and deadline violations of running instances has been studied extensively. Indicators available in event logs can be exploited to make predictions about time-related risks\cite{van2011time}. Verenich et. al. \cite{verenich2019survey} have done an extensive survey of various methods used for predicting the remaining cycle time and present a cross-benchmark comparison. Van Dongen et. al. \cite{van2008cycle} apply non-parametric regression on event data to predict remaining cycle time. Similarly \cite{evermann2017predicting} employ RNN to predict the duration of activities. Similarly, Tax. et al.\cite{tax2017predictive} propose a prediction method based on Long Short-Term Memory (LSTM) neural networks.    

\subsection{Process Outcome Predictions}

The problem of case outcome prediction aims at identifying process instances that will end up in an undesirable state (measured as likelihood and severity of fault occurrence or violation of compliance rules) \cite{verenich2016general}. Instances can be labelled as normal or deviant and then process related risks can be classified using any of the traditional modern machine learning techniques. The case outcomes can be assessed primarily by first checking if the process has met its 'hard goals' and then soft goals (determined by KPIs such as time quality, cost etc.) Similarly, in Compliance monitoring techniques are aimed at preventing Compliance violations by monitoring ongoing executions of a process and checking if they comply with respect to certain business constraints\cite{verenich2016general}. 

\subsection{Discussion}

\textbf{\textit{Explainability}} \textit{Explainablility} or \textit{Interpretability} remains a key challenge in predictive business process monitoring. \textit{Interpretability} is defined as \textit{"the ability to explain or to present in understandable terms to a human"} and often stems from incompleteness in problem formulation leading to unquantified bias \cite{doshi2017towards}.  Business process stakeholders and decision makers can only fully trust prescriptive support systems that offer an explanation for the decisions or recommendations made \cite{adadi2018peeking}. Decision Support systems which leverage machine learning methods must therefore be able to explain the reasoning behind certain decisions, recommendations, predictions made or actions taken. Predictive process monitoring methods should explain why the predictive model was mistaken when the predictions are inaccurate.  By making systems and models interpretable we want to ensure that decisions and data can be explained to end users in a transparent and easy to understand manner \cite{adadi2018peeking}. 
Increasingly, \textit{Explainability and Interpretability} is not just a desirable property but increasingly becoming a serious matter of public debate. In some countries compliance laws will require \textit{'a right to explanation'} which means end-users can ask for an explanation of a certain decision that affects their lives \cite{goodman2017european}.  This also means that use of black box methods (such as Deep Neural Networks) in predictive and prescriptive approaches will be infeasible as we can't explain the decisions or predictions made. \textit{'Explainablility'} and \textit{'establishing trust'} in these models remains one of the key challenges, not only for sensitive domains like healthcare but now for everyday consumer-facing products as well. Therefore, developing systems that provide trustworthy explanations and the necessary chain of reasoning that led to particular decisions and outcomes remains a significant challenge for future research work in predictive and prescriptive monitoring. Several methods have been proposed to make systems more comprehensible. Rizzi et. al. \cite{rizzi2020explainability} for example propose a method that uses post-hoc explainers and encoding for identifying the most common features that explains incorrect predictions. By reducing the impact of identified features, explanations can be used to improve model accuracy. Brunk et. al. \cite{brunk2020cause} take a different approach to the problem of making transparent context-sensitive predictions by proposing a next event prediction technique based on dynamic Bayesian network. Galanti et. al. \cite{galanti2020explainable} propose a based framework based on game theory of Shapley Values for explaining the predictions. Their framework is based  based on LSTM models, and can explain any generic KPI. Lastly, Verenich et. al. \cite{verenich2019predicting} employ flow analysis technique for predicting quantitative performance indicators of running process instances. Their technique can be used to estimate values of this performance indicator (e.g cycle time)   by aggregating performance indicators of the activities composing the process. We refer the readers to the work by  Doshi-Velez and  Kim \cite{doshi2017towards} and Adadi and Berrada \cite{adadi2018peeking} for a more detailed discussion around the topics of \textit{'Interpretability'} and \textit{'Explainable AI'}.

%Similarly, In \cite{breuker2016comprehensible}. Breuker et. al. have discussed a promising probabilistic modeling technique (based on grammatical-inference) to predict behaviour of running process instances.
\section{Optimizing Process Decisions through Goal-Directed Decision Support}

Business process management assists organizations in planning and executing activities that collectively deliver business value, usually in the form of a product or a service. Flexible execution of business process instances entails multiple critical decisions, involving various actors and objects, taken to achieve optimal process outcomes\cite{teinemaa2019outcome}.These decisions are of variable nature and context dependent as Business operate under uncertain real-world environments. Sub-optimal decisions during process execution, such as picking the wrong execution path, incorrect resource allocation, can affect business processes outcomes leading to cost overruns and missed deadlines \cite{ghattas2014improving}. These decisions therefore require careful attention, therefore, ability to guide and automate decision making, therefore, is crucial to maintaining and improving business process performance\cite{groger2014prescriptive}.  Overall, Analytics-driven decision support for process users and knowledge workers remains a significant challenge for BPM research \cite{catalkaya2013enriching}. 

%. e.g resource allocation, selection of appropriate paths in Control-flow, and decisions embedded in process activities\cite{ghattas2014improving} that have an affect Business processes outcomes.  
% with a view to maintaining and improving business process performance.   The 

Historically, Process support has largely focused on development of Workflow Management Systems and later evolving to Business Process Management Systems(BPMSs). Such systems have played a significant role in facilitating analysis, improvement, and enactment of business processes\cite{schonenberg2008supporting}. However, future systems will tackle the challenge of leveraging historic and current contextual data for providing intelligent assistance to process users, and in guiding process-related decisions. We foresee future process analytic decision-making systems or decision-support would take the role of: 

\begin{itemize}

\item Recommending the best suffix for a task sequence, which, if executed, will lead to a desired outcomes or desired performance characteristics. This means identify a process path that would yield best performance in a given context.

\item Providing support for Knowledge-intensive processes to assist knowledge workers. e.g. Generating optimal action recommendations for a process instance while accounting for various sources of uncertainty. 

\item Providing support for risk-aware Process Management. e.g. by early identification of process associated risks and generating recommendations for corrective actions that can help avoid a predicted metric deviation. 

\item Providing support for Resource Management. e.g. assess the suitability of a resource in executing a certain task and support resource allocation decisions by suggesting optimal work assignment policies 

\item Providing strategic support for robust process execution and for developing robust strategic plans in adversarial settings while carefully balancing multiple objectives
  
\end{itemize}

We discuss some of these ideas in detail below: 

\subsection{Risk-Informed Decision Making} 

Modern organizations operate in dynamic environments and face the challenge of dealing with uncertainties(risks) associated with various process decisions due to environmental factors that are often not fully under their control. Managing process-related risks by making risk-informed decisions remains a key challenge for organizations.  Risk reduction usually involves decreasing the likelihood and severity of faults during process execution and ensuring that desired performance goals are met. 

Management of risks during process execution remains a challenging problem\cite{suriadi2014current}. The ability to detect potential metric deviations early on is valuable in giving process owners decision information for a timely intervention.  
One primary goal of process decision support during Business process execution is to guide risk-informed decisions at run-time by modeling, detecting and mitigating risks as early as possible. In practice, this means using predictive and prescriptive techniques to identify process instances that are likely to get delayed or to terminate abnormally. This involves using those predictions to intervene early, recommend actions or support resource allocations decisions at run-time that enable organizations to avoid or decrease the likelihood of metric overrun.

To tackle these challenges, Risk-based decision support (RDS)(Sometimes known as Risk-aware Business Process Management Systems) have been proposed to help with early detection of factors by predicting risks in terms of metrics deviations during process execution. They can make recommendations that can help minimize the predicted process risk and avoid negative outcomes.  The performance criteria of business processes are typically described by Key Performance Indicators (KPIs), and their target values can be defined based on business goals.  They can assist in risk management by monitoring KPIs, PPMs, and QoS metrics. Overall, reduce risks (in terms of metrics deviations during process execution) to provide decision support for a given process, e.g., recommending the next process activity, which minimizes process risks. When those deviations exceed a tolerance threshold, interventions can be taken to decrease the likelihood of unfavourable outcomes. 

Risk reduction techniques allow us to identify process instances that are at risk of not meeting certain performance criteria and recommend preventive actions to process participants. Several techniques have been proposed in the literature for modeling and detection of risk.  Suriadi et. al. \cite{suriadi2014current}. Provide a comprehensive review of techniques for managing process-related risks.  Literature on prescriptive business process monitoring consists of techniques \cite{groger2014prescriptive} \cite{conforti2013supporting} \cite{schonenberg2008supporting} that can be used to recommend  preventive actions in order to support  \emph{risk-informed decision making}. 

Future Business Process Management systems must, therefore, assist process users/stakeholders in risk-informed decision making by generating predictions and recommendations that help reduce such risks \cite{di2017eye}. Risk Management and BPM were historically separate fields, and their integration is understudied, leaving room for future research contributions \cite{suriadi2014current}. Conforti et al. \cite{conforti2015recommendation}  discuss the Risk-aware BPM lifecycle where each phase can be complemented with elements of risk management like Risk Identification, Risk-aware Execution and Risk monitoring etc. Lastly, problems like real-time risk detection, resource scheduling, Automated risk mitigation and Real-Time Risk Monitoring are also investigated under the umbrellas of risk-informed decision making\cite{conforti2011history} \cite{conforti2012automated} \cite{conforti2013real}. Overall, managing process-related risks by making risk-informed decisions remains a key challenge for organizations.

%TODO: 
%https://www.sciencedirect.com/science/article/abs/pii/S002002551630144X

\subsection{Supporting business process execution via Process-Aware Recommender Systems}

In the previous section, we discussed how predictive monitoring techniques can provide operational support based on models learned using historic logs in order to predict what is likely going to happen next and use those prediction capabilities to influences the outcomes of running case instances. e.g. monitor processes and issue recommendations to workers and managers based on probability that a given case will violate the set performance targets. The output of Predictive business process monitoring techniques, is just \emph{predictions}. Predictions can be used as \emph{early warnings} for taking risk informed decisions but do not explicitly support answering of question like \emph{What action should we take next to achieve a particular goal?} and \emph{Why should we do it?}\cite{LEPENIOTI202057}. Compared to descriptive and predictive business analytics, prescriptive process analytics remains less mature \cite{eili2021systematic}. Marquez et. al. \cite{marquez2017predictive} point out that  \emph{`little attention has been given to providing recommendations'}. Instead of providing specific action recommendations, literature on business process monitoring focuses on forecasting future process events(and outcomes) while leaving the action implementation part to the subjective judgment of process users and business decision makers\cite{dees2019if}.

Recommender Systems have found applications in information filtering system such as in video or music services as playlist generators or content recommendations for social media companies etc\cite{beheshti2020towards}. 

\textit{Process-aware Recommender Systems} have been proposed to assist knowledge workers, in operational decision-making by recommending actions for executing a particular process/task, manage resource allocation policies and so on \cite{beheshti2020towards} \cite{schonenberg2008supporting}. Eili et al. \cite{eili2021systematic} provide a  systematic review of Recommender Systems in Process Mining and classify recommendation approaches as \emph{`pattern optimization'}, \emph{`risk minimization'}, or \emph{`metric-based'}. For structured processes, \textit{Process-aware Recommender Systems} have been proposed to assist knowledge workers, in a \textit{context-aware} adaptable fashion by recommending actions for executing a particular process/task, manage resource allocation policies and so on \cite{beheshti2020towards} \cite{schonenberg2008supporting}.
Such systems leverage technologies like machine learning to build recommender systems that monitor process instances, predict future process states and recommend appropriate actions\cite{dees2019if}.

We should note that most process-aware recommender techniques focus on supporting \emph{risk-informed decision making}.Their major focus is on preventive measures early warning recommendations to for example avoid predicted metric deviation. Another major goal of Process-aware Recommender systems 
should be to support decisions that 
that maximize the likelihood of achieving business goals. e.g Weinzier et al. \cite{weinzierl2020prescriptive} consider problem of recommending \emph{next best actions} that lead to optimal outcomes. Their technique relies on explicitly adding control-flow knowledge to their proposed technique via formal process model and uses process simulations to verify and filter the predictions of the trained predictive model. Groger et al. \cite{groger2014prescriptive} introduce the concept of recommendation-based business process optimization to support adaptive process execution. Their framework recommends actions for the next process step to take for a given process instance. For organizations data-driven process optimization enables decision support for real-time process optimization. e.g. shortening the reaction time of decision-makers to events that may affect changes in process performance.

\subsection{Operational decision support for Knowledge Intensive Processes }

Business Processes assist organizations in organizing activities that deliver business value, usually in the form of a product or a service.  Over the last decade, automation has caused the landscape of work to change significantly and Knowledge workers are now regarded as the most valuable organizational assets.  Knowledge work is characterized by unstructured processes which can be hard to specify at design-time  Supporting knowledge workers involved in the execution of unstructured Knowledge-Intensive Processes by providing context-specific recommendations remains an interesting challenge.

Knowledge Intensive Processes (KIPs) are processes that require precise expert(tacit) knowledge, involvement of knowledge workers, and consisting of activities that do not have the same level of repeatability as structured processes\cite{di2015knowledge}. Instead of assuming a rigid process structure, knowledge-intensive processes (KiPs) are goal-oriented, often unstructured(with pre-defined fragments) and characterized by activities that cannot be anticipated(or modeled in advance). KIPs represent a shift from the traditional process management view (where process models are structured with repeatable tasks), to a model where task execution depends on knowledge workers as primary process participants \cite{di2015knowledge}.   

Knowledge workers are highly trained and have specialized expertise in performing complex tasks autonomously and are considered a key asset for modern businesses. Supporting knowledge work when rigid definitions of process models are not available or cannot be designed apriori (with structured or unstructured data)remains a key challenge\cite{di2012knowledge}. They typically rely on their experience based intuition and domain expertise, for decision-making. Their work is less characterized by explicit procedures and more by creative thinking that usually cannot be planned a priori\cite{di2015knowledge}. An example of knowledge work is Clinical decision-making for patient treatment in the hospital emergency room, which is highly case-specific and requires a knowledge-driven approach. i.e Treatment Decisions are made based on highly specific medical domain knowledge, the context in the form of patient's medical history, years of specialized experience and evidence that emerges from patient test results and real-time sensors. 

Knowledge workers still lack the adequate decision support tools to assist them in executing knowledge-intensive processes\cite{di2015knowledge}. As we enter the knowledge economy, future process management systems will have to drive processes in modern enterprises that are highly knowledge-driven, are semi-structured or unstructured while leveraging a diverse range of process-related datasets. e.g recommend appropriate actions to knowledge workers while operating in dynamic environments. In \cite{di2015knowledge} Ciccio et al. have provided a set of requirements for process-oriented systems to support knowledge-intensive processes. One of the inherent challenges that future process management systems must address is that of flexibility. Flexibility means instance-specific adaptations based on context and environment. Flexible execution of business process instances involves multiple critical decisions at each step. e.g. what task to perform next and what resources to allocate to a task and so on. Previously, to address the problem of supporting flexible knowledge-intensive process, many paradigms have been proposed. e.g Adaptive Process Management(ACM), Flexible Process Aware Information Systems, case handling systems and declarative processes. Adaptive Case Management gained the most popularity amongst the various paradigms. 

\textbf{Adaptive Case Management(ACM)} is aimed at supporting knowledge workers involved in the execution of dynamic, unstructured knowledge intensive processes(KIPs) where course of action for the fulfillment of process goals is highly uncertain\cite{hauder2014research} \cite{motahari2013adaptive}. ACM offers a way to manage the entire life-cycle of a ``case" by following the 'planning-by-doing' principle, where work is done by considering the context and is continually adapted based on the changing characteristics of the environment\cite{motahari2013adaptive}.  In the \textit{case management} paradigm, the focus is on the case and its hard to pre-define the sequence of activities. For example, Case is the `Product' being manufactured or a `patient' being treated where primary driver of case progress is the case data and information that emerges as the case evolves. There however can exist template or patterns that represent the structured aspects of the process. Here process could be seen as a recipe for handling cases of specific type \cite{van2003case}.  A \textit{Case template} is created by the knowledge worker, allowing high degree of flexibility in executing a particular case. Templates can then be used to instantiate case instances and represents a middle ground between a completely specified structured process and an unstructured process.  Case execution allows us to gather feedback and adapt the templates to be reused in a particular context\cite{caseMang2016}. Adaptive Case Management (ACM) has been gaining significant interest for handling unpredictable situations in processes and still lacks 'common operational semantics' and a 'proper theory' \cite{hauder2014research}\cite{hewelt2016hybrid}. \\

In the context of Case Management, the problem of recommending 'next best steps' in a \textit{case management} system based on the knowledge of past similar cases(which is the focus of this work) has been addressed by Schonenberg et al. \cite{schonenberg2008supporting} and Motahari-Nezhad et al. \cite{motahari2011next}. Schonenberg et al. attempt it by first finding similar cases based on abstraction, then using support and Trace Weights to consider the relative importance of a log trace. Similarly,  Motahari-Nezhad et al. \cite{motahari2011next} have looked at the problem of decision support for guiding case resolution based on how similar cases were resolved in the past. Such support can complement knowledge workers decisions, often made using personal experience and expert knowledge\cite{di2015knowledge}. We argue that the notion of recommendations under-pins decision support for not only structured processes but across the whole spectrum of process management. i.e we see recommendations as a general-purpose mechanism for providing operational decision-support for not only structured but semi-structured and unstructured processes as well. In knowledge-intensive processes, for example, recommendations can provide intelligent assistance to process users by offering concrete support in various process related decisions like resource allocation decisions or action recommendations etc. Such form of assistance allows knowledge workers to take preemptive actions that can avoid negative outcomes \cite{schonenberg2008supporting}\cite{groger2014prescriptive}. Data-centric AI approaches hold the promise of  supporting knowledge  intensive processes and case management practices, whereby enabling flexible process execution. Khan et al. \cite{khan2021} propose a data-driven reinforcement learning based recommender system for supporting knowledge workers that considers the past execution data in addition to characteristics of the objects involved (e.g product or user). The proposed system recommends the next best action (or sequence of actions) while taking into account asset characteristics and process context.

% Case templates incorporate knowledge from experts and try to capture the best practices of knowledge work that has happened in the past.

%A typical BDI agent program consists of three components [9]: (1) A set of beliefs, which may be dynamically generated by sensor inputs. (2) A set of plans, where each plan contains a triggering event, context conditions and plan body (a set of action sequences). (3) A set of goals that an agent wants to achieve. Each goal can be achieved by executing plans in the plan library. The goals involved are related to plans recognized and are given a unique label (e.g. goal g1 can be achieved by executing plan p1). 

%We briefly review work concerning recommendation based decision support for allowing flexible process execution in the context of knowledge-intensive processes:
 
\subsection{Decision Support for Resource Management}

Resources are entities that are responsible for performing activities of a business process and must satisfy various (sometimes contradictory) business goals. Management of resources involves selecting the right resources, evaluating the efficiency of resources and optimally allocating tasks to relevant resources \cite{Renuka2018}.  Decision support for resource management provides process users, intelligent assistance on the optimal allocation of resources based on their capabilities, past performance, current workload and process characteristics.

Past execution data containing resource allocation decisions and can be leveraged to provide real-time decision support regarding the allocation of resources at an appropriate time while considering the specific context. Taking this data-driven approach improve productivity and efficiency of business processes and helps avoid performance deviations. Selecting the most appropriate resource and assigning it to the right task is often a challenging decision as it is dependent on task complexity, task priority and expertise of the worker. Traditional resource allocation decisions are made based on profile matching(perceived capabilities), which often involves human judgment. Such approaches are not always optimal because of various challenges such as resource unavailability, overloading and uncertainty(of process execution and resource behaviors) \cite{sindhgatta2015learning}. Traditional thinking also held that the performance of a process is determined by its design; thus, well-designed
careflows would lead to better patient outcomes. More generally, though, process context also plays an important role. Process context can be defined as knowledge exogenous to a process, and not consumed as input to a process that nevertheless serves as a determinant of process performance and proposed a context-aware recommender system for identifying context and using it to support resource allocation and task allocation decisions. 

A number of methods related to learning, reasoning, and planning resource allocation have been proposed in the literature. e.g. In \cite{Renuka2018} Rajan et al. have explored addressed the question of
context-aware process management. Such recommender systems can derive data-driven business process provisioning that supports effective dispatching and staffing policies and assist managers in meeting the desired quality of service (or performance) levels. Russel et al. \cite{russell2005workflow} discuss workflow resource patterns and have identified three main allocation types, namely, capability-based allocation(where we match capabilities of available resources with and requirements of an activity), history-based allocation (using past execution data to make ) and finally Role-based allocation(which considers the organizational position and relation of the resource). Liu et al. \cite{liu2014q} model the task allocation problem as a Markov Decision Processes (MDPs) and show their Q-Learning based method overcomes many of the shortcomings of traditional methods(e.g. load imbalance) to compute social relation between two resources.  We can also consider various abstraction levels when allocating resources. e.g. Arias et al. \cite{arias2016framework} have proposed a recommendation system that dynamically allocates resources at a sub-process level based on multi-factor criteria (to assess resources). Their proposed tool considers metric scores in the various dimensions of the resource process cube(knowledge base) to present a ranked list of suitable resources.  Similarly, machine learning approaches can be used to mine resource allocation rules. e.g. Huang et. al. \cite{huang2011reinforcement} treat resource allocation as a sequential decision making optimization problem and propose a reinforcement learning based, resource allocation solution. Their Q-learning based framework allows adjustment of real-time allocation decisions by learning appropriate allocation policies based on available feedback.

%An associated challenge is that of knowledge management as in today's competitive business landscape, knowledge especially those that knowledge workers posses is considered a key strategic asset by the organisations. 

\subsection{Strategic Decision Modeling}

%Decision Support at Strategic Level (or via Goal Modeling)}

{\textit{Goal-orchestration for flexible process execution: }} Goal models holds the promise of delivering significant value  Strategic Modelling, by providing a hierarchic representation of statements of stakeholder intent, with goals higher in the hierarchy (parent goals) related to goals lower in the hierarchy (sub-goals) via AND- or OR-refinement links. Goal models encode important knowledge about feasible, available alternatives for realizing stakeholder intent represented at varying levels of abstraction. A number of prominent frameworks leverage goal models, including KAOS, i* and Tropos \cite{1319986}. Strategic Modelling for organizations selecting amongst alternative goal refinements (OR-refinements). Given a goal model that delimits that space of goals and sub-goals that an organization can seek to satisfy, this is a critical (and indeed, only) decision problem to be solved. An AND-refinement of a goal is a statement of know-how that tells the organization how to achieve a parent goal (although without sequencing information, and thus falling short of being a full procedure or process model). OR-refinements offer alternative specifications of know-how for a given parent goal. 

Replacing tasks or activities with goals in-process models allows us to enact processes in, flexible, context-sensitive ways. In \cite{santipuri2017goal} Santipuri et al. introduce the concept of goal orchestration for modeling processes (or process behavior) to enable flexible process execution.  Goal orchestrations offer abstract, strategy-level views on processes, which can aid human understanding and ease process redesign. Their proposed technique can mine goal orchestrations from enterprise event logs and compute alternative task-level realizations of a goal if the initial attempt at realizing the goal fails to achieve the desired results. Goal-oriented process mining is a promising sub-field. In \cite{ghasemi2020event} Ghasemi et al. present a survey various techniques for  mining goals from event logs and argue that combining goals and process mining can potentially augment the precision, rationality and interpretability of mined models.

%https://www.ipvs.uni-stuttgart.de/departments/as/downloadgallery_as/groegech/Groeger_Prescriptive_Analytics.pdf
 %Instead of following a pre-determined path, adaptive process management systems can make

\textit{Achieving resilience in adversarial settings:} The need to future-proof businesses is widely acknowledged as one of the hardest challenges facing business decision-makers. Businesses need to anticipate environmental changes and the likely behavior of competitors. Much of what happens in the business environment (the  effects  of  moves  by  these  actors) is adversarial in nature and adversarial moves prevent or impede the achievement of business goals. Decision-making in adversarial settings involves reasoning about chains of moves and counter-moves by the adversarial entities involved. Strategic resilience requires that businesses make decisions that are most resilient to adversarial moves by players in the business environment. e.g Red teaming allows businesses to informally reason about the adversary's strategic planning process\cite{hoffman2017red}. Similarly, in \cite{gou2017leveraging} Gou et al. present a decision support framework for robust process enactment. They leverage adversarial game search (e.g Monte Carlo game tree search) to compute alternative flows in order to anticipate and account for possible ways in which the execution environment might impede a process from achieving its desired effects or outcomes. This notion can be extended for strategic decision-making as well, where decision-making is viewed through an abstraction of a two-player game. Here  goal models (representing structured models of strategy) can be combined with game-tree search using augmented game trees to assist organizations in selecting amongst alternative goal refinements (OR-refinements). The final computational machinery, overall, provides business management with a resilient strategic decision-making framework that can reason the consequences of various decisions.

%In \cite{} Gou et al. present a Game-theoretic notion where decision-making is viewed through an abstraction of a two player game. They combine goal models(representing structured models of strategy) with game-tree search using augmented game trees to assist organizations in selecting amongst alternative goal refinements (OR-refinements). The proposed computational machinery, overall, provides business management with a resilient strategic decision-making framework that can reason the consequences of various decisions. 

\section{Discussion and Future Directions}

In this section, we present an overarching picture of some of the key challenges and characteristics needed to develop future process support systems. We can also view these characteristics as feature requirements for building future business process management and decision support systems.  We argue, that in addition to supporting the existing capabilities, the next generation of business process management systems will offer several additional analytics features such as process monitoring, resource allocation, risk management. Some of these have already been incorporated by existing vendor service and product offerings (e.g. Process Monitoring support in Apromore), while problems like agility support for unstructured knowledge-intensive processes still remain unaddressed. 

%Data-centric AI approaches hold the promise of supporting knowledge-intensive processes as retrospective data embodies rich experiences, representing expert decisions and associated outcomes \cite{di2015knowledge}. We have seen in the previous sections, how such data can be leveraged to build predictive and prescriptive models which enable organizations to improve decision making in dynamic environments. 

In \cite{dumas2022augmented} Dumas et. al present a vision of AI-Augmented Business Process Management Systems. In such systems execution flows are not pre-determined rather use AI technology to adapt and reason within  a set of restrictions based on one or more performance indicators. This allows operates largely autonomously, within the boundaries set by the process frame. Keeping this in mind, We briefly discuss enabling AI technologies for supporting flexible execution of processes and attain the pre-defined goals of a given business process. One useful abstraction is to see Processes as a Sequential decision-making activity, where we must make a sequence of decisions in response to information about the outcomes of our actions as we proceed. Decisions can cover the management of given resources, interventions for achieving process goals, and supporting decisions or the kind discussed in this section. The problem gets further complicated when processes are deployed in stochastic environments, where the outcomes of our actions are uncertain. There exist many approaches for designing decision-making systems and the problem of \textit{``learning and decision making over time to achieve a goal''} has been studied by multiple disciplines and they all provide interesting perspectives \cite{sutton2022quest}. From an agent perspective, There are many methods for designing decision-making agents. They differ in the
responsibilities of the designer and the tasks left to automation. We provide a brief overview of methods that can be applied to tackle sequential decision problems faced in process analytics.

%selected and promising (and less explored) technique for building process analytics system that exhibit reward-driven Behaviour and can leverage such date to provide decision support systems: 

%Techniques like Planning, behvaiour cloning via Machine Learning and Reinforcement Learning hold the potential for building such systems. 

\textit{Reinforcement Learning} \cite{sutton2018reinforcement} provides a framework for learning from interaction with the environment in order to achieve a goal (implicitly defined by the reward function). \textit{Reinforcement Learning} has widely been used to model sequential decision problems and has shown great promise in solving large scale complex problems with long time horizons, partial observability, and high dimensionality of observation and action spaces\cite{berner2019dota}. 

Reinforcement Learning (RL) assumes that there is an agent operating in the real world. At each step $t$ the agent, Executes action $A_{t}$, Receives an observation $O_{t}$ and Receives scalar reward $R_{t}$. The Problem can be formulated as a Markov Decision Process\cite{sutton2018reinforcement} defined by $(\mathcal{S}, \mathcal{A}, T, R)$ tuples where $\mathcal{S}$ and $\mathcal{A}$ refer to the state and action spaces; $T: \mathcal{S} \times$ $\mathcal{A} \rightarrow \mathcal{S} $, is the state transition function and $R: \mathcal{S} \times \mathcal{A} \rightarrow \mathbb{R}$ represents the reward function. The goal of the agent is to estimate an optimal policy $\pi: \mathcal{S} \rightarrow \mathcal{A}$  or an optimal action value function $ q_{\pi}(s, a)=\mathbb{E}_{\pi}\left[G_{t} \mid S_{t}=s, A_{t}=a\right]$  which maximizes the expected return  $
\mathbb{E}\left[\sum_{t=1}^{L} \gamma^{t} R_{t} \mid \pi\right]$ over a given MDP\cite{sutton2018reinforcement}. 

\textit{Reinforcement Learning} can allow us to formulate process goals as the maximization of a cumulative reward which is a very powerful general-purpose idea. In process analytics, it can be applied to for example provide decision support in the form of interventions or action recommendations which are used to generate possibilities from which human workers can pick the best alternative given the additional context and experience they have access to. i.e RL allows us for general formulation of sequential decision problems under the assumption that the model is known and that the environment is fully observable.

\textit{Offline Reinforcement Learning} in particular provides an excellent opportunity to build \textit{adaptive systems} that learn from past experience and leverage feedback in the form of rewards. We can use   \textit{offline Reinforcement Learning} to learn the decision criteria representing optimal outcomes and use it to recommend optimal action based on the current state of the process. \textit{Offline Reinforcement Learning}  requires sufficiently diverse training data that is close to cases that system might encounter in the future. Further it requires that we define a \textit{reward function} (manifested as the weighted soft goal score), which  captures the goals and priorities of the specific process. A \textit{Reinforcement Learning} based decision support system takes a state based view for supporting process related decisions. Such systems can consider context and recommend actions in each stage of the process(which assumes availability of effect log). \\

\noindent
\textit{Building adaptive systems via BDI agents:} optimal decision making in a sequential context requires reasoning about future sequences of actions and observations. In \textit{Artificial Intelligence}, a \textit{rational agent} will perceive its environment, use its internal knowledge base, along with reasoning and planning capabilities, to select actions that lead to the desired goal state(or close) according to some utility measure\cite{russell2002artificial}. \textit{Agent-oriented} programming deals with modelling and writing software systems with this concept of rational agent, in which each component, or agent, perceives the environment through sensors and acts on the environment with actuators\cite{shoham1993agent}. The \textit{Belief-Desire-Intention (BDI)} agent is a particularly popular and effective architecture for designing such agents. A typical \textit{BDI agent} program consists of three components \cite{rao1995bdi}: First, a set of beliefs acquired potentially through sensor inputs. Secondly, a set of plans, where each plan has an associated triggering event, pre-defined context conditions and plan body consisting of a set of action sequences. Lastly, a set of goals that an agent wants to achieve. Goals are related to plans. i.e. Each goal can be achieved by executing plans in the plan library.  There are many programming languages and platforms developed over the last few decades for implementing BDI-agent systems. Some of these languages include PRS (Procedural
Reasoning System) \cite{ingrand1992architecture}, AgentSpeak(L) \cite{rao1996agentspeak}, Jack [25], dMARS (Distributed Multiagent Reasoning System) etc. 

Mining agent programs allows organisation to quickly build agent programs that can potentially replace traditional software systems which are  costly and hard to maintain. In \cite{xu2013automatic} Xu et. al. propose a framework for learning BPI agent plans from process and effect logs. The authors propose a \textit{plan recognition framework} for generating \textit{BDI style plans}. The framework requires input in the form of behaviour logs generated by enterprise applications in order to mine a \textit{`draft'} version of agent code that can potentially replace some of the applications deployed inside the organisation. Specifically, a \textit{WF-Net} is generated using ProM using a number of pre-defined transformation rules and then the generated \textit{WF-Net} is transformed into a set of plans using the proposed algorithm. Lastly, effects logs are used to identify context which becomes the pre-condition for each of the extracted plan. The plans utilise both positive and negative example sets and uses norm learning mechanisms to infer normative plans.

\textit{Robotic Process Automation:} Firms are interested in the identification of potential areas of automation to save costs and improve efficiency.  Historically Business Process Management Systems (BPMSs) supported Business process automation (BPA) by executing process instances, supporting the distribution of work to process participants and delegating activities to various information systems deployed across the organization (e.g. checking the creditworthiness of an applicant) \cite{dumas2018business}. 

Today we observe a rise of new technologies that can enable Business Process Automation by automating procedural work and supporting complex processes. Promising technologies like Robotic Process Automation(RPA) and Reinforcement Learning hold the promise of replacing human worker and automate repeated tasks. Task automation might mean replacing human workers entirely with intelligent agents, while decision automation means making decisions that humans previously made. In other scenarios, it facilitates organizations to automate human decision making. e.g. automated allocation of resources in complex knowledge-intensive scenarios might mean providing decision support(e.g. in the form of recommendations) for resources involved in process execution.  

RPA represents an interesting shift, aiming to automate parts of business processes that consist of humans interacting with day-to-day software(e.g. transferring data from an Enterprise Resource Management system to a web application form). \textit{``Robotic Process Automation (RPA) is an emerging technology that allows organizations automating repetitive clerical tasks by executing scripts that encode sequences of fine-grained interactions with Web and desktop applications''}.  Alternatively  \textit{``RPA is an umbrella term for tools that operate on the user interface of other computer systems in the way a human would do''}

In the context of Process Analytics, RPA aims at automating business processes that consist of human interaction with software and provide decision support for resources involved in process execution. Implementing RPA is suitable in situations where processes that are too infrequent for traditional process automation to be profitable, but still repetitive  enough to be formalized into an RPA process mode. We can identify opportunities of autommation from logs of interactions between workers and Web and desktop applications. Frameworks like Value-driven RPA \cite{kirchmer2019value} are useful identify right sub-processes to automate, given the process context. 

%In robotic process mining, we can discover routines from user interactions.  

\textit{Automated Process Improvement:} In traditional process mining, process optimization occurs as a result of post-mortem data analysis. In Automated Process Improvement, we strive for proactive improvement of business processes during process execution and attempt to automate as many aspects of a given process as possible. Automated Process Improvement is one of the most desired capability in process Analytics and can make processes more adaptive \cite{zur2015business}. Here we are interested in an intelligent exploration of improvement strategies using domain knowledge and all available historical and current process-related data(generated by enterprise systems and sensory data). 

Search-based optimization techniques allow us to identify and discover opportunities, for improving business processes, from event lots. We do so by considering various performance metrics. e.g the cycle time and process cost as key performance indicators(KPIs). Automated Process Improvement can identify various opportunities of improvement. It can also be used to streamline tasks execution where we identify control-flow related improvement opportunities. e.g re-ordering, merging and parallelization of tasks. It can also identify opportunities for task automation e.g Using RPA. Secondly, we can identify best Practices and give recommendations based on the analysis of best performing instances in the past. Similarly, Automated Process Improvement can also help with optimal resource allocation. As discussed earlier, we use historic data to come with up optimal resource allocation policies. For example, we can optimally design staff schedules by analyzing historical data and using process goals formulated as Key Performance Indicators and systematically evaluate the proposed changes. Lastly, we can optimize decision logic to improve the routing of the cases, which means adding or removing decision points or enhancing existing decision rules. 

Gröger et. al. \cite{groger2014prescriptive} introduce the concept of recommendation-based business process optimization (rBPO). Such a system can generate action recommendations during process execution, enabling us to perform process optimization using pre-specified metrics. Apart from mining based approaches, predictive methods can generate action recommendations during process execution. Lastly, Bozorgi et. al. \cite{dasht2020process} show that treatment recommendations (based on causal machine learning), when applied during the execution of a case can improve the overall outcome of a process.

\section{Conclusion}
Modern organizations routinely deploy process analytics, including process discovery and variant analysis techniques, both to gain insight into the reality of their operational processes and also to identify process improvement opportunities. 
Process analytics refers to the repertoire of techniques centred on process mining, predictive monitoring, decision and automation support. Process analytic approaches play a critical role in supporting the practice of Business Process Management and continuous process improvement by leveraging process-related data to identify performance bottlenecks, reducing costs, extracting insights, and optimizing the utilization of available resources. They also
enable us to mine insights from process data (which encompasses process logs but include many other types of
data as well), predict the behavior of process instances and provide operational and strategic decision support.  In this work, we briefly surveyed  the literature on process analytics and identified promising directions for future research.

%\begin{acks}
%To Robert, for the bagels and explaining CMYK and color spaces.
%\end{acks}

\bibliographystyle{ACM-Reference-Format}
\bibliography{main}

\appendix

\end{document}